\documentclass{article}

\usepackage{graphicx} 
\usepackage{authblk}
\usepackage{url}
\usepackage{enumitem}
\usepackage{booktabs, tabularx, makecell}
\usepackage[most]{tcolorbox}
\tcbuselibrary{listings, breakable}
\usepackage[top=2.5cm, bottom=2.5cm, left=2cm, right=2cm]{geometry}
\usepackage{threeparttable}

\newcommand{\para}[1]{\textbf{#1.}\quad}

\title{The backbone of science: analysis of citation networks between papers and their sources}

\author[1]{Wonhee Jeong}
\author[2]{Dimitri Marinelli}
\author[3]{Satyaki Sikdar}
\author[1] {Gaurang Singh Yadav}
\author[1, *]{Santo Fortunato}
\affil[1]{Center for Complex Networks and Systems Research (CNetS), Indiana University, Bloomington, IN, USA}
\affil[2]{Institute of Complex Systems (UBICS) and Department of Condensed Matter Physics, University of Barcelona, Barcelona, Spain}
\affil[3]{Department of Computer Science, Loyola University Chicago, Chicago, IL, USA}
\affil[*]{corresponding author: santo@iu.edu}

\date{\today}

\begin{document}

\maketitle

\begin{abstract}

The bibliography of scientific papers lists items with variable degree of relevance for the contents of the paper itself. 
If we could identify the sources, i.e., the works that actually inspired the paper, their citations
can help us uncover the genesis of scientific projects
and would be more representative of the actual importance of papers and authors than the standard citation counts, when all references are considered. Here we present an analysis of the \textit{backbone of science}, i.e., the network of citations between papers and their sources. The latter are extracted from the full body of papers via Large Language Models (LLMs), which are currently very capable of correctly identifying the context in which a paper is cited. Using two different but related prompts, we find that the LLMs select only a small set of references, not taken at random, and that the resulting backbone networks are quite similar to each other with respect to their in-degree distributions, modularity, transitivity, and degree correlations. Backbone networks have higher heterogeneity in their in-degree distributions, compared to the full network, but the most cited papers are usually the same, with some important exceptions. Citation rankings among authors are also remarkably stable. We conclude that the full citation network, despite its redundancy with respect to the backbones, presents a reliable picture of the relative citation impact of papers and authors.

\end{abstract}

\section{Introduction}

The list of references of a scientific paper is supposed to include all previous works that, in one way or another, are closely linked to it. Some may be general references, like review articles and books, that summarize the state of the field of the paper. Most of the others are usually related to the specific contents of the paper, which may have used an idea, problem formulation, or a method introduced in them. Not all references are equally important for the genesis of the paper, though. We call \textit{sources} the building blocks of the focal paper, i.e., those references which provided a key idea or methodology, without which the paper would not have been possible.   
Let us suppose that we identified all sources of all papers. The resulting citation network could be considered the \textit{backbone of science}, in that it only includes the papers providing the main ingredients of all others. Besides its intrinsic value, the backbone of science allows us to derive a more reliable estimate of the impact of papers and authors than the full citation network, which entails a lot of redundancy due to the many references having only a marginal role in the conception and design of the papers.

In this work, we compare the full citation network with its backbone. A key question is how to identify the sources, and how to do it at scale, for thousands of papers. For this, we used Large Language Models (LLMs)~\cite{zhao2026survey}, which are increasingly adopted by scholars in their citation practices~\cite{zhang23}. Most of the times LLMs are used to find relevant citations for a given work~\cite{algaba2025deep,algaba2025large}, or even specific paragraphs~\cite{press24}, though hallucinations might lead to incorrect references~\cite{mugaanyi2024evaluation}. Also, one can use LLMs to help scholars screen citations~\cite{oami2025optimal}. 

Here we query LLMs with prompts that specifically request to output a selection of relevant references. We used two different, albeit similar, prompts to mitigate the risk that results are too biased by the specific adopted query. While we did not put any restrictions on the number of selected references, the resulting lists often include a handful of items, so the resulting backbone networks are much sparser than the full citation network. 

Overall, we find that the two backbones are structurally similar to each other, even if the overlap between references selected by the LLMs using the two prompts is low. Citation distributions are skewed and with similar exponents across all three networks, with the backbones displaying higher heterogeneity. Interestingly, rankings of papers and authors according to their number of citations in the full network are quite similar to the corresponding rankings in both backbones, suggesting that the relative importance of papers is not biased by the much larger number of citations of the full network.

\section{Results}

\subsection{Characterization of Extraction Selectivity and Backbone Networks}\label{sec:Jaccard}

We used \textit{DeepSeek-R1-Distill-Llama-70B} to extract references from papers of five scientific fields: Network Science, Quantum Computing, Gravitational Waves, Stochastic Finance, and Natural Language Processing (Section~\ref{m:data}).
Each field yields a citation network $G$, from which we derive two backbone networks, $H$ and $I$, via the selected references, using two distinct prompts (Section~\ref{m:LLM}).

The edges of these backbone networks are supposed to represent the key citations of a paper.
According to Table~\ref{tab:net_info}, backbone networks exhibit a significantly lower number of edges compared to the full citation networks. Despite the absence of explicit constraints on the number of references selected by the LLM (detailed in Section~\ref{m:LLM}), backbone networks preserve only approximately~$15\%$ to~$25\%$ of the original citations.

To quantify the overlap of the extracted reference sets, we calculate the Jaccard similarity index ($J$)—defined as the ratio of the intersection of two sets to their union—for each paper and average the results over all papers~\cite{REF_Jaccard}.
Since only few references are selected, a full comparison would feature multiple cases in which there is either full or no overlap. For this reason, we restricted the comparison to papers for which at least one query returned 3 references or more.
According to Table~\ref{tab:Jaccard}, low $\langle J(H, I)\rangle$ values reflect the fact that the LLM selects rather different subsets of references from different prompts. 

\begin{table}[ht]
\centering
\caption{Summary statistics of the full citation network ($G$) and respective backbone networks ($H$ and $I$), for five different scientific fields: (a) Network Science, (b) Quantum Computing, (c) Gravitational Waves, (d) Stochastic Finance, and (e) Natural Language Processing. While the number of nodes ($N$) remains invariant, the number of edges ($E$) varies based on how many references are extracted by the LLM in each case, with percentages in parentheses calculated relative to $E(G)$.}
\label{tab:net_info}
\begin{tabular}{lrrcc}
\hline \hline
   & $N$ & $E(G)$ & $E(H)$ & $E(I)$ \\ \hline
Network Science      & $15\,804$  &  $76\,160$   &   $15\,596$ ($20.5$\%)  &  $14\,928$ ($19.6$\%)   \\
Quantum Computing    & $18\,274$  &  $79\,670$    &   $13\,131$ ($16.5$\%)  &  $12\,971$ ($16.3$\%)   \\
Gravitational Waves  & $15\,379$  &  $77\,141$    &   $16\,628$ ($21.6$\%)  &  $16\,077$ ($20.8$\%)   \\
Stochastic Finance   & $9\,715$  &  $9\,699$   &   $2\,298$ ($23.7$\%)  &  $2\,633$ ($27.1$\%) \\
Natural Lang. Proc.  & $6\,990$  &  $22\,711$   &   $4\,515$ ($19.9$\%)  &  $4\,786$ ($21.1$\%)   \\
\hline \hline
\end{tabular}
\end{table}

\begin{table}[ht]
\centering
\caption{Average Jaccard similarity indices $\langle J(H,I) \rangle$ between the lists of references of each paper extracted by the two prompts, restricted to cases where at least one of the prompts selected more than 3 references, for the five fields we examined. The standard error of the mean is indicated in parentheses.
}
\label{tab:Jaccard}
\begin{tabular}{lr}
\hline \hline
    &  $\langle J(H, I)\rangle$ \\ \hline
Network Science &  $0.302~(6)$  \\
Quantum Computing & $0.337~(6)$  \\
Gravitational Waves & $0.255~(6)$  \\
Stochastic Finance & $0.350~(3)$  \\
Natural Lang. Proc. & $0.300~(1)$  \\
\hline \hline
\end{tabular}
\end{table}

\begin{figure}[ht]
\centering
\includegraphics[angle=0,width=1\columnwidth]{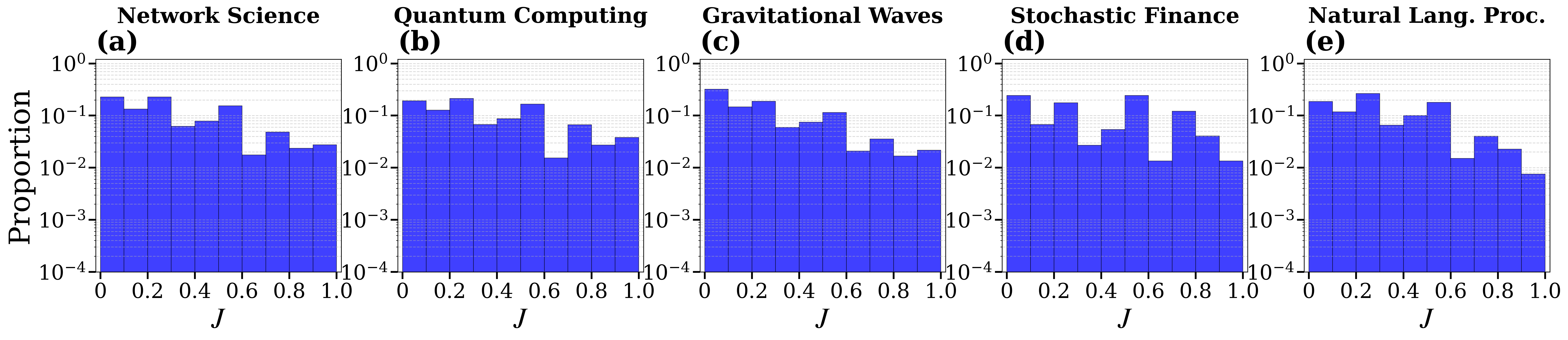}
\caption{Histogram of the Jaccard similarity index $J(H, I)$ between lists of references selected by the two different prompts across the five fields, restricted to cases where at least one of the prompts selected more than 3 references. The histograms utilize 10 bins. Table~\ref{tab:Jaccard} provides the corresponding average values for each distribution.}
\label{fig:Jaccard}
\end{figure}

\subsection{In-degree Distributions}\label{sec:indegree}

Our first goal is to analyze and compare the structure of the three networks $G$, $H$ and $I$ for each field. We start with deriving their in-degree distributions. Since the in-degree $k$ of a paper in a citation network is the number of citations of the paper, the in-degree distributions $P(k)$ correspond to the citation distributions of the same set of papers in the three settings. For each backbone network we also generate two baselines, to check whether the LLM selects the references randomly, or according to their number of citations (Section~\ref{m:network}).

\begin{figure}[ht]
\centering
\includegraphics[angle=0,width=1\columnwidth]{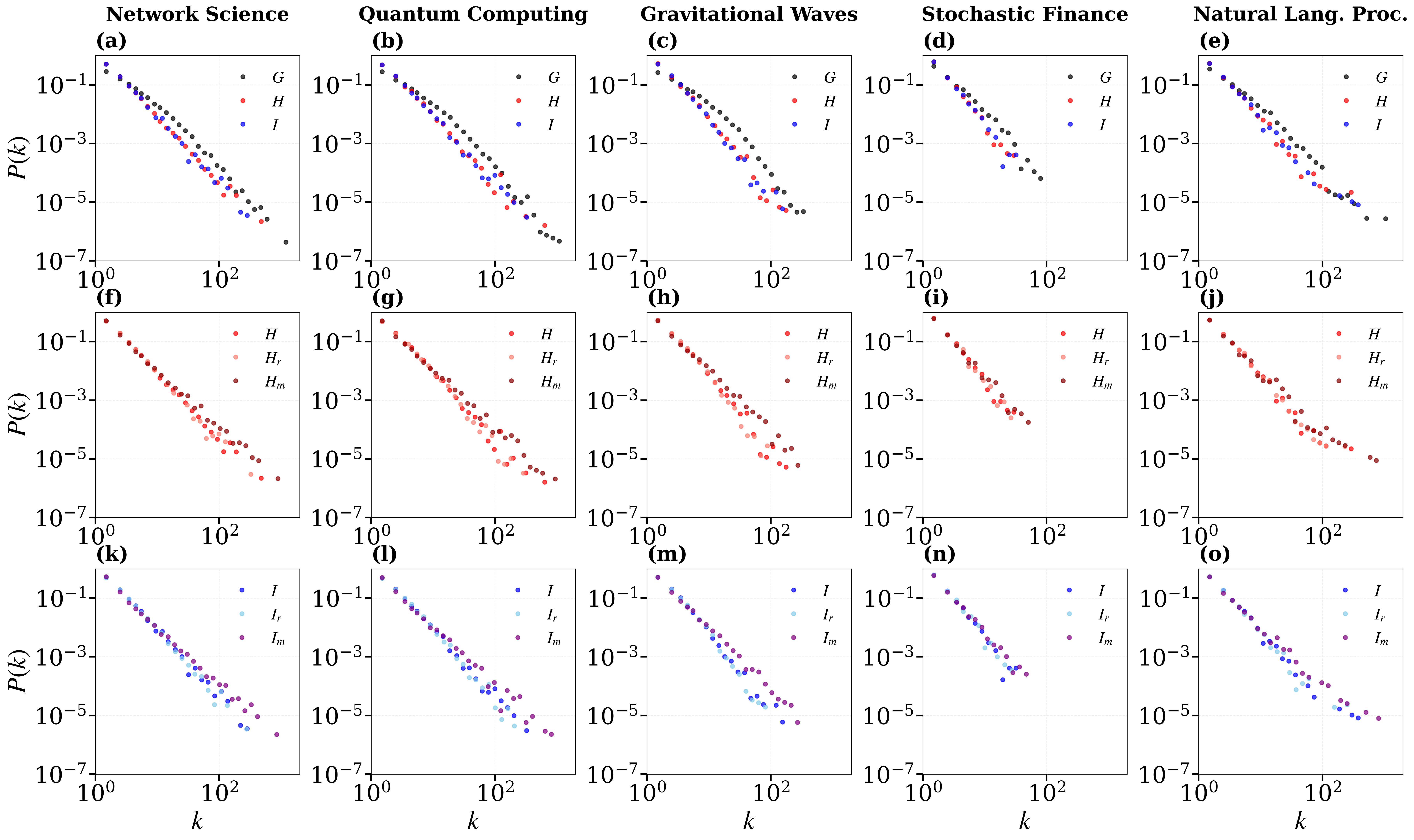}
\caption{
In-degree distributions across research fields and network types. Panels (a–o) are organized in a $3 \times 5$ grid where columns represent research fields: Network Science (a, f, k), Quantum Computing (b, g, l), Gravitational Waves (c, h, m), Stochastic Finance (d, i, n), and Natural Lang. Proc. (e, j, o). 
Rows indicate specific comparison types: the top row (a–e) compares the full-citation network ($G$, black) with the two backbone networks, $H$ (red) and $I$ (blue); the middle row (f–j) illustrates $H$ alongside its baselines, $H_r$ (salmon) and $H_m$ (dark red); and the bottom row (k–o) displays $I$ and its baselines, $I_r$ (sky blue) and $I_m$ (purple).
}
\label{fig:degree}
\end{figure}

As shown in Fig.~\ref{fig:degree}, all networks exhibit fat-tailed in-degree distributions, a well-known feature of citation networks~\cite{price76,radicchi08}. The plots in the top row show that both the full citation networks and the backbone ones have power law in-degree distributions on their tails. For the backbone networks the whole distribution follows power law patterns~\cite{REF_powerlaw}. 
Table~\ref{tab:exponent} reveals that the power-law exponents ($\gamma$) for $G$ are higher than those for the backbone networks (with the exception of Stochastic Finance). A lower exponent in the backbone networks indicates a heavier tail, which means that the LLM-driven extraction process favors a set of highly cited, and possibly influential, papers.

To better understand the LLM-driven reference extraction process, we
compared each backbone network with two baselines that simulate two different ways to select the same number of references. In the first one (\textit{most-cited}) we assume that references are selected based on their number of citations, which generates the networks $H_m$ and $I_m$. In the second one (\textit{random}) we assume that references are selected randomly, which generates the networks $H_r$ and $I_r$. Details are provided in Section~\ref{m:network}. The middle and bottom rows of Fig.~\ref{fig:degree} illustrate the comparisons of the in-degree distributions of $H$ and $I$ with their respective baselines. As summarized in Table~\ref{tab:exponent}, the most-cited baselines yield lower exponents than the original backbone networks, due to the artificial selection of highly cited papers. The random baselines conversely distribute citations across a broader set of papers, resulting in similar or higher exponents. 

The distinguishability of the distributions is evaluated using the two-sample Kolmogorov-Smirnov (KS) test~\cite{twoKS}. Results are reported in Table~\ref{tab:KS}. As expected, the full-citation network is statistically distinguishable from the backbone networks, and the most-cited baselines are also distinguishable from their original backbone networks.
On the other hand, the random baselines often appear indistinguishable from the original backbone networks. Notably, the distributions of $H$ and $I$ are found to be statistically indistinguishable from each other.  

We also quantify the correlation of nodes' in-degrees using Pearson and Spearman correlation coefficients~\cite{REF_Pearson,REF_Spearman}.
According to Tables~\ref{tab:Pearson} and~\ref{tab:Spearman}, correlation scores are very high throughout.
Spearman coefficients are generally lower than the Pearson values, which is primarily due to the heavy-tailed nature of the distributions, leading to a high frequency of tied ranks among the low-degree nodes.
Despite being constructed from smaller subsets of base references, backbone networks exhibit high correlation and a similar in-degree distribution to the full-citation network. Remarkably, this strong consistency holds as well for $H$ and $I$, as well as with their respective baselines.

To provide a concrete illustration of this structural consistency at the top of the hierarchy, Table~\ref{tab:rank} lists the highest in-degree papers in the Network Science field. We observe a strong overlap among the top-ranked papers across $G$, $H$, and $I$, confirming that the backbone networks preserve the relative importance of the most prominent papers. Analogous tables for the other research fields are provided in Appendix~\ref{appendix_rank}.
Therefore, while the two prompts often identify different references (Section~\ref{sec:Jaccard}), the respective backbone networks display high consistency with respect to in-degree.

\begin{table}[ht]
\centering
\caption{
Power-law exponents ($\gamma$) of the network in-degree distributions.
This table summarizes the estimated numerical values of the exponents (with the standard error of the maximum likelihood estimate in parentheses). The corresponding in-degree distribution plots are presented in Fig~\ref{fig:degree}.
}
\label{tab:exponent}
\begin{tabular}{lrrrrrrr}
\hline \hline
    & $\gamma(G)$ & $\gamma(H)$ & $\gamma(I)$ & $\gamma(H_r)$ & $\gamma(I_r)$ & $\gamma(H_m)$ & $\gamma(I_m)$ \\ \hline
Network Science& $2.33(5)$ & $2.15(3)$ & $2.18(3)$ & $2.18(3)$ & $2.41(6)$ & $1.88(3)$ & $1.84(2)$ \\
Quantum Computing & $2.47(6)$ & $2.32(5)$ & $2.17(3)$ & $2.32(4)$ & $2.49(6)$ & $1.86(3)$ & $1.83(2)$ \\
Gravitational Waves & $3.10(1)$ & $2.51(6)$ & $2.68(8)$ & $3.00(1)$ & $2.69(5)$ & $2.14(5)$ & $1.88(2)$ \\ 
Stochastic Finance & $2.25(4)$ & $2.60(1)$ & $2.70(1)$ & $2.60(1)$ & $3.00(1)$ & $2.13(6)$ & $2.11(5)$ \\ 
Natural Lang. Proc. & $2.33(6)$ & $2.25(6)$ & $2.25(6)$ & $2.27(6)$ & $2.25(5)$ & $1.87(5)$ & $1.84(4)$ \\ 
\hline \hline
\end{tabular}
\end{table}

\begin{table}[ht]
\centering
\caption{
Results of the two-sample Kolmogorov–Smirnov test comparing the in-degree distributions.
The values indicate the KS statistic $D$ and statistical significance is denoted by asterisks based on the $p$-values ($^{*}p < 0.05$, $^{**}p < 0.01$, $^{***}p < 0.001$).
}
\label{tab:KS}
\begin{tabular}{llllllll}
\hline \hline
    & $D(G, H)$ & $D(G, I)$ & $D(H, I)$ & $D(H, H_r)$ & $D(H, H_m)$ & $D(I, I_r)$ & $D(I, I_m)$ \\ \hline
Network Science& $0.4271^{\mathrm{***}}$ & $0.4338^{\mathrm{***}}$ & $0.0108$ & $0.0093$ & $0.0503^{\mathrm{***}}$ & $0.0060$ & $0.0529^{\mathrm{***}}$ \\
Quantum Computing & $0.4461^{\mathrm{***}}$ & $0.4496^{\mathrm{***}}$ & $0.0046$ & $0.0181^{\mathrm{*}}$ & $0.0531^{\mathrm{***}}$ & $0.0120$ & $0.0567^{\mathrm{***}}$ \\
Gravitational Waves & $0.3931^{\mathrm{***}}$ & $0.3885^{\mathrm{***}}$ & $0.0079$ & $0.0171^{\mathrm{**}}$ & $0.0590^{\mathrm{***}}$ & $0.0123$ & $0.0628^{\mathrm{***}}$ \\ 
Stochastic Finance & $0.3019^{\mathrm{***}}$ & $0.2749^{\mathrm{***}}$ & $0.0269^{\mathrm{**}}$ & $0.0019$ & $0.0135$ & $0.0025$ & $0.0143$ \\ 
Natural Lang. Proc. & $0.3865^{\mathrm{***}}$ & $0.3733^{\mathrm{***}}$ & $0.0184$ & $0.0048$ & $0.0470^{\mathrm{***}}$ & $0.0074$ & $0.0519^{\mathrm{***}}$ \\ 
\hline \hline
\end{tabular}
\end{table}

\begin{table}[ht]
\centering
\caption{
Pearson correlation coefficients ($\rho_{\mathrm{P}}$) of in-degree (number of citations) between the full citation network, backbone networks, and their baselines.
This metric quantifies the linear correlation between in-degrees.
Overall, the correlation coefficients exhibit consistently high values across all networks.
}
\label{tab:Pearson}
\begin{tabular}{llllllll}
\hline \hline
    & $\rho_{\mathrm{P}}(G, H)$ & $\rho_{\mathrm{P}}(G, I)$ & $\rho_{\mathrm{P}}(H, I)$ & $\rho_{\mathrm{P}}(H, H_r)$ & $\rho_{\mathrm{P}}(H, H_m)$ & $\rho_{\mathrm{P}}(I, I_r)$ & $\rho_{\mathrm{P}}(I, I_m)$ \\ \hline
Network Science & $0.9410$ & $0.9539$ & $0.9378$ & $0.9608$ & $0.9557$ & $0.9547$ & $0.9290$ \\
Quantum Computing & $0.9121$ & $0.9319$ & $0.8601$ & $0.9255$ & $0.9644$ & $0.9405$ & $0.9174$ \\
Gravitational Waves & $0.8687$ & $0.8707$ & $0.8936$ & $0.9019$ & $0.9289$ & $0.8940$ & $0.9165$ \\ 
Stochastic Finance & $0.8783$ & $0.8793$ & $0.9164$ & $0.9217$ & $0.9011$ & $0.9297$ & $0.9202$ \\
Natural Lang. Proc. & $0.9335$ & $0.9495$ & $0.9347$ & $0.9568$ & $0.9040$ & $0.9605$ & $0.9637$ \\ 
\hline \hline
\end{tabular}
\end{table}

\begin{table}[ht]
\centering
\caption{Spearman rank correlation coefficients ($\rho_{\mathrm{S}}$) for in-degree (citation) rankings between the full citation network, backbone networks, and their baselines.
This metric quantifies the monotonic relationship between the in-degree sequences of the networks. The scores are fairly high across the board.
}
\label{tab:Spearman}
\begin{tabular}{llllllll}
\hline \hline
    & $\rho_{\mathrm{S}}(G, H)$ & $\rho_{\mathrm{S}}(G, I)$ & $\rho_{\mathrm{S}}(H, I)$ & $\rho_{\mathrm{S}}(H, H_r)$ & $\rho_{\mathrm{S}}(H, H_m)$ & $\rho_{\mathrm{S}}(I, I_r)$ & $\rho_{\mathrm{S}}(I, I_m)$ \\ \hline
Network Science & $0.7194$ & $0.7159$ & $0.7338$ & $0.7613$ & $0.7047$ & $0.6695$ & $0.5950$ \\
Quantum Computing & $0.7290$ & $0.7321$ & $0.7493$ & $0.7605$ & $0.7099$ & $0.7629$ & $0.7110$ \\
Gravitational Waves & $0.6678$ & $0.6893$ & $0.6684$ & $0.7070$ & $0.6994$ & $0.7279$ & $0.7096$ \\ 
Stochastic Finance & $0.6701$ & $0.7045$ & $0.7168$ & $0.8321$ & $0.8210$ & $0.6892$ & $0.6608$ \\ 
Natural Lang. Proc. & $0.6850$ & $0.6979$ & $0.7188$ & $0.7204$ & $0.6551$ & $0.7152$ & $0.6436$ \\ 
\hline \hline
\end{tabular}
\end{table}

\begin{table}[ht]
\centering
\caption{Comparison of the top 10 papers in the Network Science, ranked by in-degree in $G$, $H$, and $I$. The table comprises the union of the top 10 lists from each network and is sorted by the in-degree in $G$, $k(G)$.}
\label{tab:rank}
\begin{threeparttable}
\small
\begin{tabularx}{\textwidth}{X c rrr}
\hline \hline

  Title\tnote{*}  & Year & $k(G)$ (Rank) & $k(H)$ (Rank) & $k(I)$ (Rank)  \\ \hline
  The structure and function of complex networks & 2003 & $1\,334$ (1) &$535$ (1) & $316$ (2) \\
  Emergence of scaling in random networks & 1999 & $1\,081$ (2) & $400$ (2) & $442$ (1) \\
  Statistical mechanics of complex networks & 2002 & $666$ (3) & $307$ (3) & $201$ (4) \\  
  Epidemic spreading in scale-free networks & 2000 & $630$ (4) & $193$ (6) & $148$ (5) \\
  Finding and evaluating community structure in networks & 2003 & $559$ (5) & $194$ (5) & $210$ (3) \\
  Community detection in graphs & 2010 & $552$ (6) & $212$ (4) & $130$ (7) \\
  Assortative mixing in networks & 2002 & $550$ (7) & $122$ (10+) & $105$ (10+) \\
  Fast unfolding of communities in large networks & 2008 & $484$ (8) & $67$ (10+) & $146$ (6) \\
  Random graphs with arbitrary degree distrib. and their apps. & 2000 & $476$ (9) & $94$ (10+) & $120$ (9) \\
  Community structure in social and biological networks & 2001 & $470$ (10) & $161$ (10) & $104$ (10+) \\
  
  Error and attack tolerance of complex networks & 2000 & $443$ (10+) & $164$ (8) & $97$ (10+) \\
  Statistical physics of social dynamics & 2007 & $442$ (10+) & $174$ (7) & $118$ (10) \\
  Catastrophic Cascade of Failures in Interdependent Networks & 2010 & $364$ (10+) & $162$ (9) & $121$ (8) \\

\hline \hline
\end{tabularx}
\begin{tablenotes}
  \small
  \item[*] Our dataset missed a handful of key papers. Therefore we added them manually to the dataset. 
\end{tablenotes}
\end{threeparttable}
\end{table}

\subsection{Structural Analysis of Networks}\label{sec:undirected}

We now investigate the connectivity patterns and structural properties inherent in the relationships between papers. 
We treat all edges as undirected, allowing us to examine the networks' mesoscale and local structure properties through three metrics: the degree-dependent clustering coefficient ($C(k)$)~\cite{PRE_Ck1, PRE_Ck2}, the average normalized degree of nearest-neighbors ($\tilde{k}_{nn}$)~\cite{pastor01}, and robustness modularity ($R$)~\cite{REF_modular}.

\begin{figure}[ht]
\centering
\includegraphics[angle=0,width=1\columnwidth]{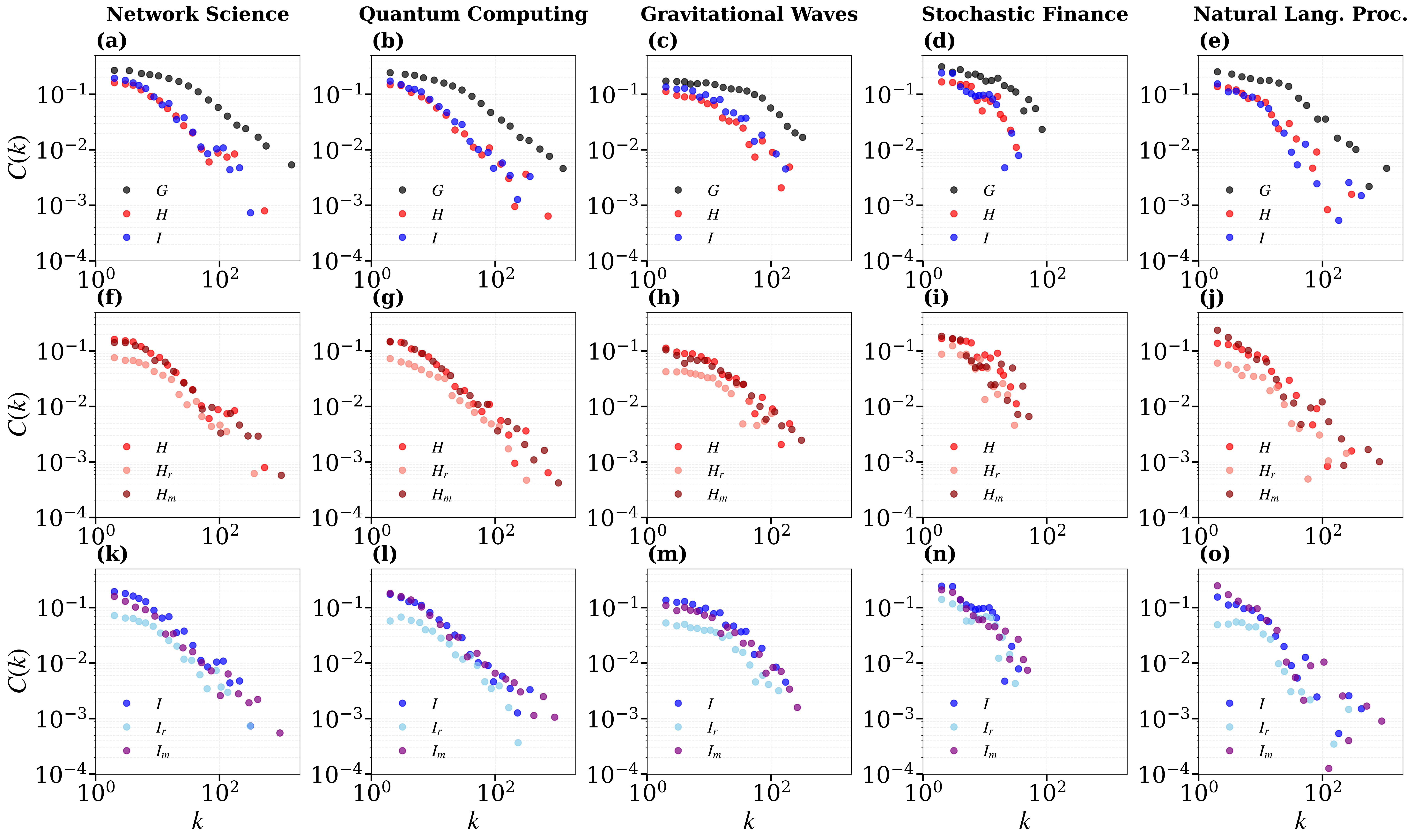}
\caption{
Degree-dependent clustering coefficient.
The layout, color scheme, and field organization are identical to those described in Fig.~\ref{fig:degree}.
}
\label{fig:Ck}
\end{figure}

\para{Local Clustering Coefficient} The local clustering coefficient measures the degree of transitivity in the neighborhood of a node as the ratio of existing edges between its neighbors to the maximum possible number of such edges. The measure can be averaged over degree classes, i.e., groups of nodes having equal degree $k$, yielding the function $C(k)$ (Figure~\ref{fig:Ck}). 

The plots in the upper row compare the full citation network ($G$) with the backbone networks ($H$, $I$). We observe that $G$ consistently exhibits higher $C(k)$ values than the backbone networks. This is expected, because
$G$ has many more edges than $H$ and $I$, leading to higher values of $C(k)$. In contrast, the patterns of $C(k)$ for the backbone networks are basically indistinguishable.

The middle row plots show the results for the backbone network $H$ and its baselines, $H_r$  and $H_m$. In the low-degree region ($k < 10$), we find that the $C(k)$ of $H$ is significantly higher than that of $H_r$. This confirms that the edges filtered by the LLM are not randomly selected. Interestingly, the $C(k)$ curves for 
$H$ and $H_m$ are very similar over the whole range of $k$-values, although these networks have different in-degree distributions (Table~\ref{tab:exponent}).
These patterns are consistently observed for the other backbone network $I$, as shown in the plots on the bottom row.

\begin{figure}[ht]
\centering
\includegraphics[angle=0,width=1\columnwidth]{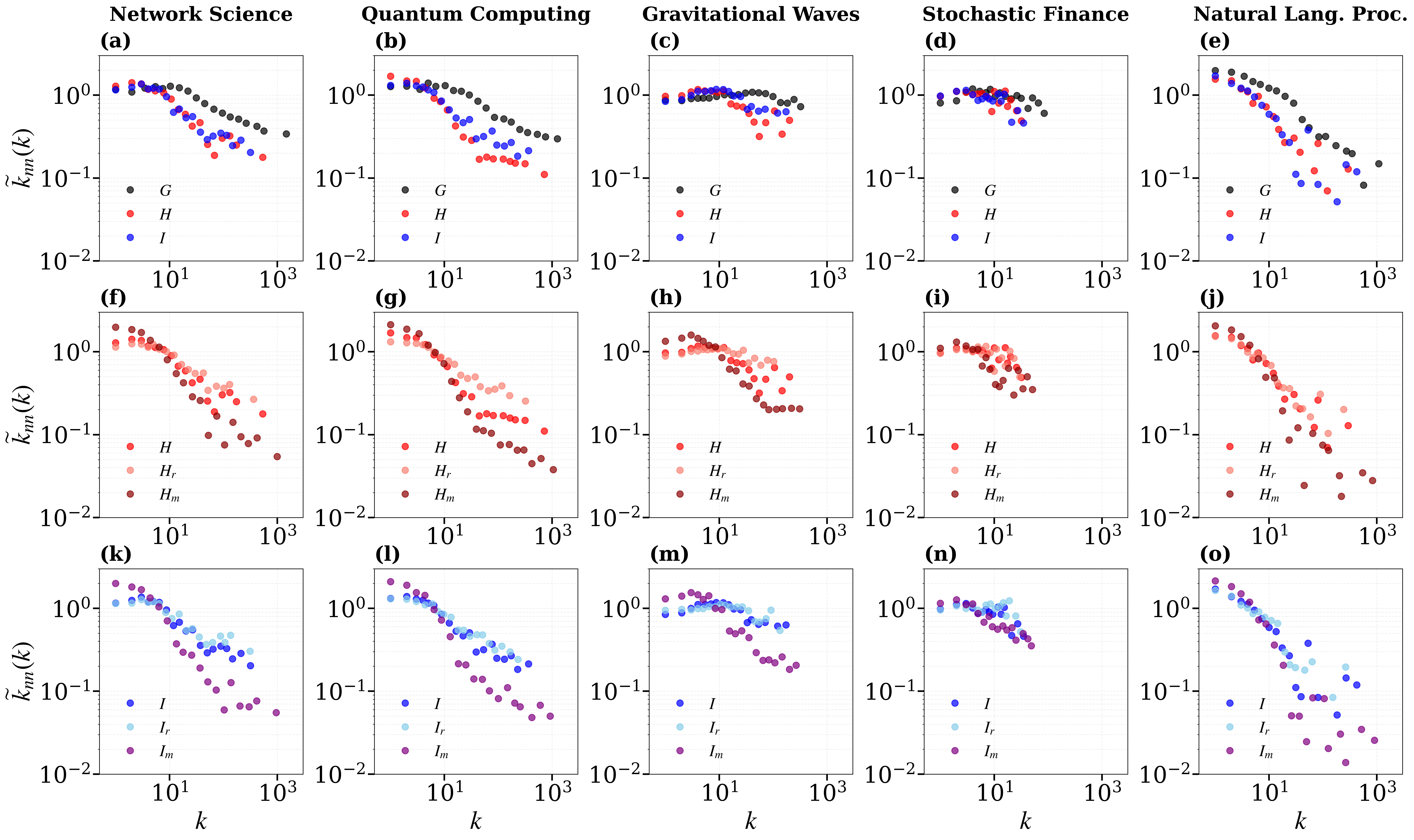}
\caption{
Average normalized degree of nearest-neighbors as a function of degree.
The layout, color scheme, and field organization are identical to those described in Fig.~\ref{fig:degree}.
}
\label{fig:knn}
\end{figure}

\para{Nearest Neighbors Degree}
Second, we examine the average normalized degree of nearest-neighbors, $\tilde{k}_{nn}(k)$, as presented in Fig.~\ref{fig:knn}. Like $C(k)$, this is also a function of $k$, obtained by averaging the mean degree of the neighbors of a node over all nodes with degree $k$, and dividing this score by the expected one obtained in the ensemble of random networks with equal degree sequence as the original one. The resulting function $\tilde{k}_{nn}(k)$ shows if the degrees of connected nodes are more or less correlated with each other than in a null model, where the edges of the network at hand are randomly repositioned: values 
above or below one indicate that the degrees of neighboring nodes are more or less correlated with each other than in the null model, respectively.

The top row compares $G$ with the backbone networks $H$ and $I$. In the high-degree regime ($k > 10$) $G$ consistently exhibits higher values of $\tilde{k}_{nn}(k)$ than the backbone networks. This indicates that the probability for a highly cited paper to cite another highly cited paper is greater in the original citation network than in the ones parsed by the LLM. 

The middle row of Fig.~\ref{fig:knn} displays the results for $H$, $H_r$, and $H_m$. 
The most-cited baseline $H_m$ shows a more disassortative trend than $H$. This is because $H_m$ is artificially biased towards high-degree (highly cited) nodes, so connections of low-degree nodes to high-degree ones are more common. The fact that $H$ shows a more moderate trend implies that the LLM does not simply prioritize highly cited papers. The other backbone network $I$ also shows qualitatively similar results, as illustrated in the bottom row plots of Fig.~\ref{fig:knn}

\para{Robustness Modularity}
Finally, we computed the robustness modularity $R$, a score that estimates the strength of the community structure of a network~\cite{REF_modular}. Results are shown in Table~\ref{tab:modular}. Across all fields, the full citation networks ($G$) exhibit the highest $R$-values. Backbone networks ($H$, $I$) show significantly higher robustness compared to their random baselines ($H_r$, $I_r$). The randomization process behind the formation of $H_r$ and $I_r$ favors a uniform distribution of edges, generating weaker communities.
Instead, the most-cited baselines ($H_m$, $I_m$) exhibit higher robustness than the backbone networks. Since the LLM preferentially selects hubs in this case, the resulting network has a cohesive community structure. 
The difference in $R$ between the backbone networks ($H$, $I$) is marginal compared to their difference from other baselines. This consistency indicates that while $H$ and $I$ are based on different prompts, they have a similar level of modular organization.

In summary, the structural analysis across multiple metrics demonstrates that the LLM-generated backbone networks are distinct from the full citation network, and their properties significantly deviate from those of basic baselines, confirming that references are selected in ways which are neither trivially random nor based exclusively on their citation scores. Nevertheless, backbone networks show remarkably similar behavior according to these metrics, showing some sort of consistency, at the population level, of the underlying selection criteria used by the LLM in response to the different prompts. 

\begin{table}[ht]
\centering
\caption{Robustness modularity ($R$) of networks.
This metric quantifies the strength of the community structure with respect to perturbations of the network structure via random edge rewiring. A value of $R \approx 0$ indicates that the network lacks significant community structure, whereas $R \approx 1$ signals very pronounced community structure.}
\label{tab:modular}
\begin{tabular}{llllllll}
\hline \hline
    & $R(G)$ & $R(H)$ & $R(I)$ & $R(H_r)$ & $R(I_r)$ & $R(H_m)$ & $R(I_m)$ \\ \hline
Network Science & $0.7997$   &  $0.3365$  &  $0.3313$ & $0.1873$ & $0.1933$ & $0.3402$ & $0.3435$ \\
Quantum Computing & $0.7884$ & $0.3245$ & $0.2996$ & $0.2564$ & $0.2431$ & $0.4188$ & $0.3861$ \\
Gravitational Waves & $0.8504$ & $0.4189$ & $0.4813$ & $0.2563$ & $0.3174$ & $0.5145$ & $0.5131$ \\ 
Stochastic Finance & $0.6752$ & $0.1347$ & $0.1567$ & $0.0810$ & $0.0689$ & $0.1688$ & $0.1744$ \\ 
Natural Lang. Proc. & $0.7218$ & $0.0790$ & $0.0759$ & $0.0524$ & $0.0597$ & $0.1620$ & $0.1752$ \\ 
\hline \hline
\end{tabular}
\end{table}

\subsection{Temporal Evolution of Networks}

So far we have focused on the static structural properties of the full citation network and backbone networks. In this subsection, we investigate their temporal evolution, to evaluate the growth of the fields.

We adopt robustness modularity ($R$) and the average in-degree ($\langle k \rangle$) of networks. For a given year $t$, we generate the cumulative networks $G^t$, $H^t$, and $I^t$. These networks include all papers and citations up to year $t$. 
Subsequently, we measure robustness modularity~\cite{REF_modular}, and the average in-degree at year $t$, which is defined as:
\begin{equation}
\langle k \rangle ^{t} = \frac{E^{t}}{N^{t}}, \label{eq:k+}
\end{equation}
where $E^t$ and $N^{t}$ represent the cumulative count of citations and papers up to year $t$, respectively.
By definition, $\langle k \rangle^{t}$ grows
if the ratio of newly added citations to newly added papers in year $t$ is larger than the average in-degree in the previous years.
Conversely, a lower marginal ratio leads to a decrease in $\langle k \rangle^t$.
Since $G$ includes all citations within a field, the temporal evolution of its $\langle k \rangle^t$ serves as a proxy for the quantitative growth of the field.
In contrast, the backbone networks consist of citations selectively filtered by the LLM, and they represent the qualitative growth of the field.

\begin{figure}[ht]
\centering
\includegraphics[angle=0,width=1\columnwidth]{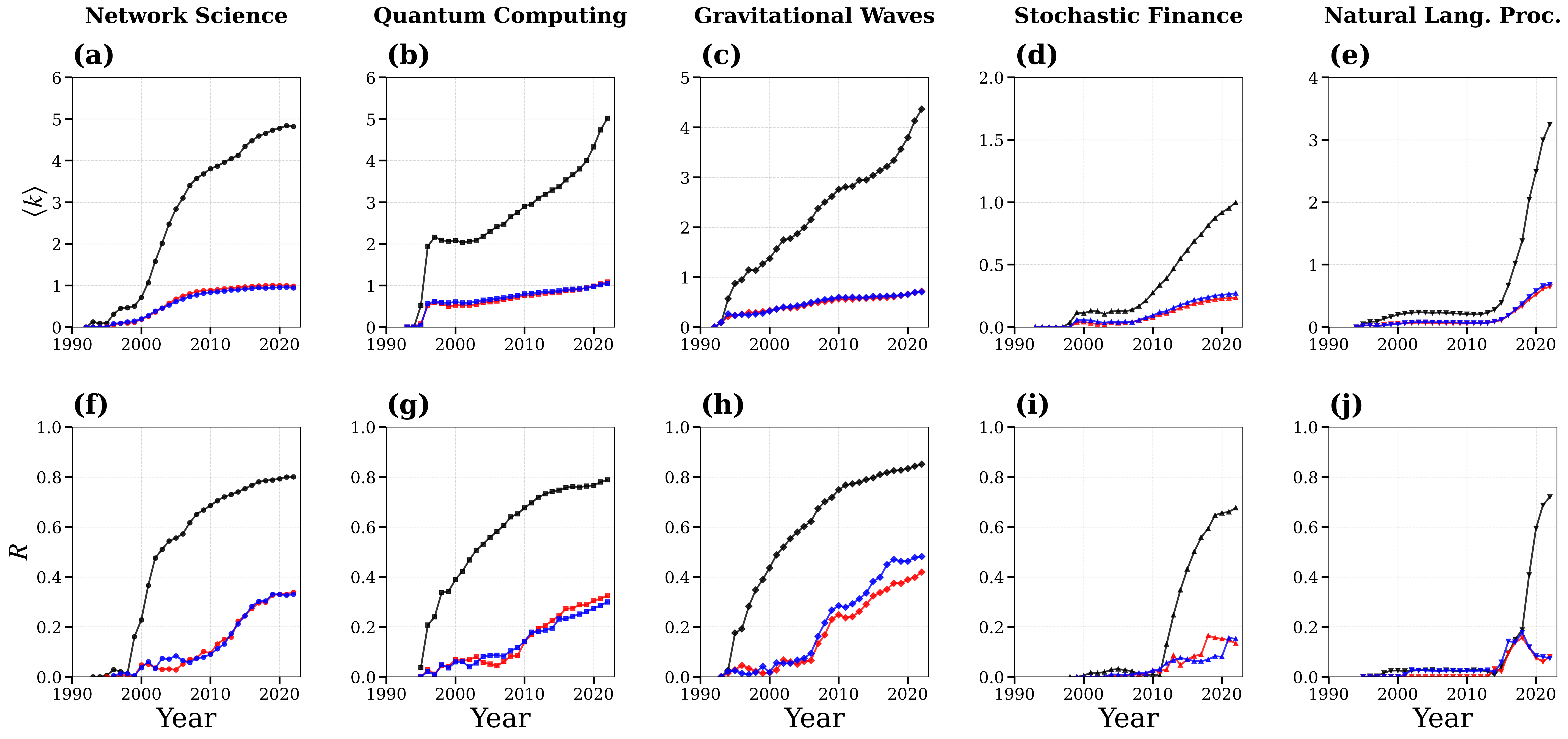}
\caption{
Temporal evolution of the average in-degree (top panels) and robustness modularity (bottom panels) across five research domains: (a, f) Network Science, (b, g) Quantum Computing, (c, h) Gravitational Waves, (d, i) Stochastic Finance, and (e, j) Natural lang. Proc.. Each field is uniquely identified by its marker shape: circles (Network Science), squares (Quantum Computing), triangles (Gravitational Waves), diamonds (Stochastic Finance), and inverted triangles (Natural Lang. Proc.). The trajectories compare the persistent growth in the full-citation network ($G$, black) with the characteristic saturation observed in backbone networks $H$ (red) and $I$ (blue). 
}
\label{fig:evolution}
\end{figure}

According to Fig.~\ref{fig:evolution}(a)-(e), in the cases of Quantum Computing and Gravitational Waves, $\langle k \rangle^t$ of the full-citation network increases rapidly, whereas for the backbone networks the growth is much slower. This suggests that, while overall the fields display a growing supply of citations, the ``good'' citations selected by the LLM are approximately constant over time.
In the cases of Network Science, Stochastic Finance, and Natural Lang. Proc., there are distinct periods during which $\langle k \rangle^t$ increases dramatically across all three networks, though the growth for the backbone networks remains markedly slower than for the original ones. 

The bottom panel of Fig.~\ref{fig:evolution} illustrates the temporal evolution of robustness modularity. Generally, the full citation networks exhibit substantially higher robustness modularity compared to their backbone networks ($H$ and $I$). Notably, in Fig.~\ref{fig:evolution}(f), (i), and (j), $R(G^t)$ experiences abrupt surges, signaling large-scale growth within the respective disciplines.
These rapid increases in $R(G^t)$ coincide with an upward trend in the average in-degree. 
In Network Science, the sharp rise around the late 1990s coincides with the seminal introductions of small-world~\cite{Nature_smallworld} and scale-free~\cite{Science_scalefree} network models, which led to the field's expansion. 
Stochastic Finance exhibits a similar surge after 2010, which coincides with the academic community's intense focus on advanced risk modeling following the 2008 global financial crisis.
In Natural Language Processing (NLP), the introduction of Word2Vec~\cite{Word2Vec} and the attention mechanism~\cite{Attention} triggered a sharp rise in $R$.
Conversely, unlike the sudden spikes seen in other areas, fields such as Quantum Computing and Gravitational Waves exhibit a consistent and gradual increase in $R(G^t)$ from the earliest years of our data. This continuous growth pattern suggests that these disciplines have experienced steady, long-term research attention.

In contrast, the backbone networks ($H$ and $I$), owing to their extreme sparsity, increase slowly and yield consistently lower $R(G^t)$ values. Mirroring the trends observed in the average in-degree, the structural robustness of $H$ and $I$ remains practically indistinguishable.
Prior to 2005, the backbone networks for Network Science, Quantum Computing, and Gravitational Waves exhibit negligible modular robustness. Thereafter, $R(G^t)$ begins a steady upward trajectory, eventually establishing a distinct modular structure by 2022.
Stochastic Finance is remarkably sparse among the analyzed fields, possessing the lowest average in-degree. Because of this structural sparsity, even though $R$ begins to rise slowly after 2010, it fails to surpass $0.2$ by 2022, signifying that the modular structures within $H$ and $I$ remain extremely fragile against random perturbations.

The evolutionary trajectory of NLP presents a striking anomaly in Fig.~\ref{fig:evolution}(j). The robustness modularity in its backbone networks reaches a maximum around 2008 and subsequently declines. Following the explosive paradigm shifts driven by the Word2Vec~\cite{Word2Vec} in 2013 and attention mechanism~\cite{Attention} in 2017, the field attracted unprecedented academic interest and experienced hyper-accelerated growth: the average in-degree increase abruptly as shown in Fig.~\ref{fig:evolution}(e).
In the context of community structure, this massive influx of new citations acts as cross-community bridges. The heavy inter-community linking effectively blurs the boundaries between distinct modules, thereby weakening the overall community structure.
While all scientific disciplines accumulate citations over time, this phenomenon of modularity reduction driven by massive attention and a sudden flood of works appears to be characteristic of the AI-related domain.

Overall, the full citation networks consistently exhibit a marked upward trend in both average in-degree and robustness modularity, which aligns with the broader academic phenomenon of ``citation inflation''~\cite{pan18,petersen19}. In contrast, backbone networks ($H$ and $I$) demonstrate significantly slower growth in both metrics, signaling that there is substantial redundancy in the full citation networks. Crucially, the temporal trajectories of $H$ and $I$ regarding both $\langle k \rangle^t$ and $R(G^t)$ are remarkably similar across all analyzed fields. This longitudinal consistency corroborates the structural similarities observed in Secs.~\ref{sec:indegree} and~\ref{sec:undirected}.

\subsection{Evaluation of Selection Intent via Excess Degree}\label{sec:significant}

In Section~\ref{sec:Jaccard}, we showed that the two prompts extract generally distinct sets of base references. However, in the subsequent sections, we showed that structural and temporal features exhibit remarkable consistency between the backbone networks, $H$ and $I$.
Since the traditional network metrics we used are insufficient to capture paper-level preferences, a new metric is required to identify which specific papers are being selected by the LLM.

We introduce the \textit{excess degree} ($S$),
representing the deviation of a paper’s in-degree within the backbone networks ($x \in \{H, I\}$) from its expected in-degree in the proportionally scaled full-citation network. 
For a given paper $i$, the excess degree $S_i$ is defined as:

\begin{equation}
S_{i}(x) = k_{i}(x) - \frac{E(x)}{E(G)} k_{i}(G) , \label{eq:significant}
\end{equation}

where $k_{i}(G)$ and $k_{i}(x)$ denote the in-degrees of paper $i$ in the full-citation and backbone network, respectively, and $E(x)$ represents the total number of edges in the backbone network.
The second term on the right-hand-side of Eq.~\ref{eq:significant} is the expected value of the degree of $i$ if one reduces the number of edges to that of $x$ by removing edges at random.
By construction, the backbone network $x$ is a subgraph of $G$, which implies the inherent constraint $k_{i}(x) \le k_{i}(G)$. Given this boundary, a positive $S_i$ indicates that the LLM ``overselects" the paper with respect to its global citation count, whereas a negative value suggests underselection.

\begin{figure}[ht]
\centering
\includegraphics[angle=0,width=1\columnwidth]{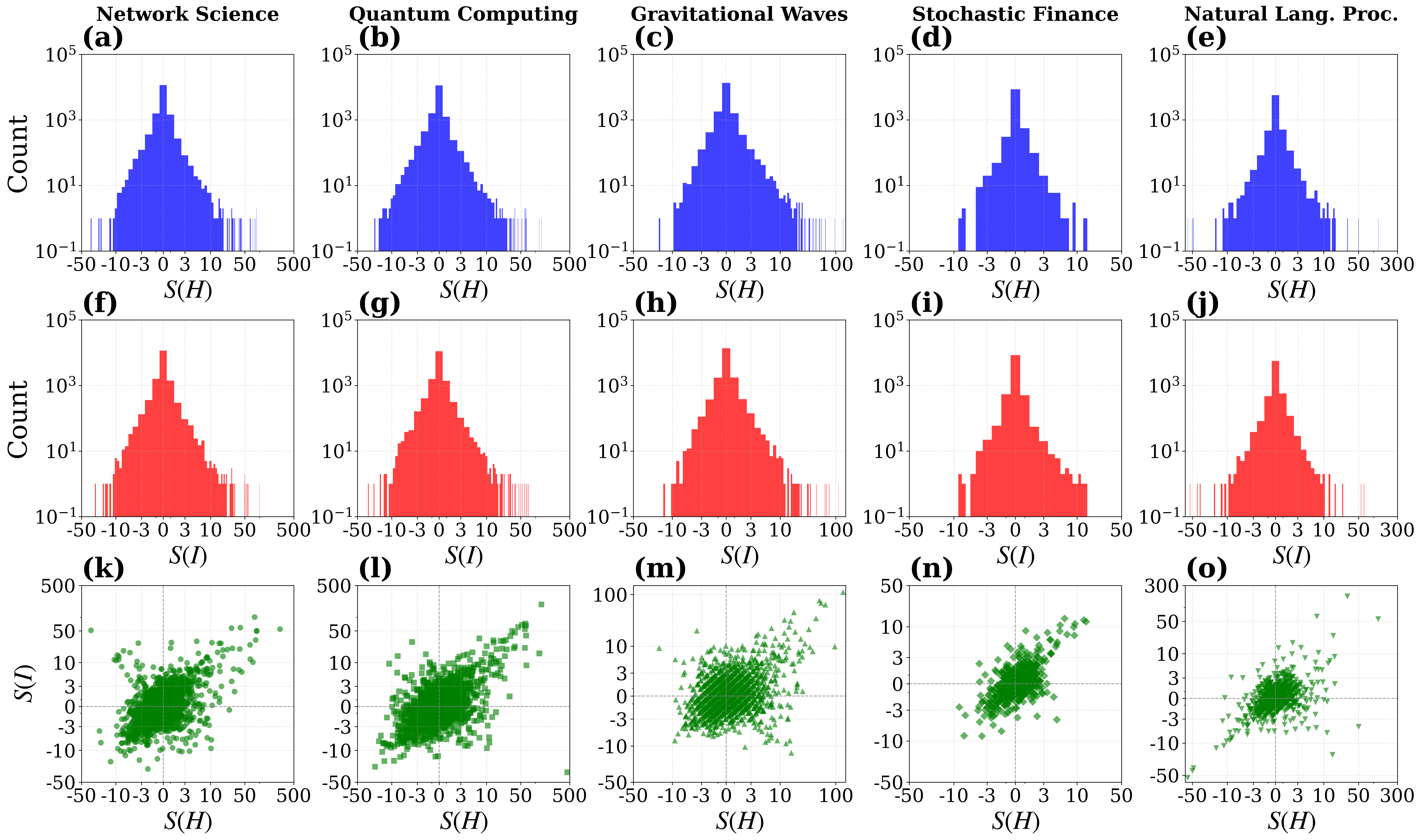}
\caption{
Node-level excess degrees ($S(H)$, $S(I)$) across five fields. 
Rows categorize the analysis type: the top row (a–e) and middle row (f–j) display histograms of excess degrees for backbone networks $H$ and $I$, respectively, while the bottom row (k–o) presents scatter plots comparing $S(H)$ against $S(I)$.
To show the full distribution of excess degree, including small values that would otherwise be obscured by extreme outliers, a symmetrical logarithmic (symlog) scale is applied. The axes maintain a linear scale within the interval $[-3, 3]$ and transition to a logarithmic scale outside this range.
}
\label{fig:significant}
\end{figure}

We evaluate the S-scores of all papers across the two backbone networks. Figure~\ref{fig:significant} presents the distributions of these scores alongside two-dimensional scatter plots in the $S(H)$--$S(I)$ plane. As shown in the histograms, the overwhelming majority of papers cluster within a narrow range between $-10$ and $10$. 
Such narrow concentration occurs because most papers in the full citation network possess very low citation counts, which inherently limits their statistical deviation and naturally confines their scores near zero.
Nevertheless, the distributions exhibit outliers, and there are subsets of papers with exceptionally high or low $S$ scores.

The scatter plots in Fig.~\ref{fig:significant} (k–o) provide a direct comparison between the $S$ scores of nodes in the backbone networks. 
If the two prompts produced very similar lists of references selection criteria, we would expect most papers to cluster closely along the identity line ($y=x$). However, the scatter plots clearly reveal a sizable number of papers positioned far from this line, indicating significant differences in the two selection criteria.

A representative example is the Louvain algorithm paper in Network Science~\cite{blondel08} ($S(H) = -32.11$ and $S(I) = 51.13$). This work has had a profound impact in the discipline by providing a highly efficient technique for community detection, enabling the rapid identification of community structures in large-scale networks. Its starkly negative $S(H)$ but exceptionally high $S(I)$ score reflects that it is frequently cited as a foundational methodological tool or in applications, rather than as a general background reference. This case illustrates that the LLM’s selection is not a simple measure of popularity, but rather a context-aware process that follows the specific intent of the prompt.

The node-level heterogeneity observed above is quantitatively confirmed by the correlation analysis. 
As summarized in Table~\ref{tab:correl_sign}, both the Pearson and Spearman correlation coefficients for the excess degree on the two different backbone networks are markedly lower than those observed for in-degree in Tables~\ref{tab:Pearson} and~\ref{tab:Spearman}.
While the backbone networks share hubs, the relative importance assigned to these nodes varies significantly across prompts.

Based on these results, excess degree reveals variations at the node level that were previously unobserved by using traditional network metrics. These fine-grained distinctions highlight whether specific papers are selected by the LLM in response to the unique criteria of each prompt.
As such, these findings demonstrate that our methodology identifies a specific scientific backbone corresponding to the qualitative requirements of each prompt, confirming that the LLM's selection is driven more by the context than by the simple citation counts.

\begin{table}[ht]
\centering
\caption{
Pearson and Spearman correlation coefficients of excess degree between backbone networks.
While the coefficients remain positive, they exhibit lower values compared to those of in-degree correlations (Tabs.~\ref{tab:Pearson} and~\ref{tab:Spearman}).
This trend reflects the nature of the excess degree as a metric for relative paper importance; the inclusion of negative values increases the sensitivity of the metric to prompt variations, leading to more pronounced fluctuations between the two backbones.
}
\label{tab:correl_sign}
\begin{tabular}{lll}
\hline \hline
    & $\rho_{\mathrm{P}}(H, I)$ & $\rho_{\mathrm{S}}(H, I)$ \\ \hline
Network Science & $0.5653$ & $0.3917$ \\
Quantum Computing & $0.3438$ & $0.4315$ \\
Gravitational Waves & $0.6587$ & $0.3527$ \\ 
Stochastic Finance & $0.6424$ & $0.4047$ \\ 
Natural Lang. Proc. & $0.5384$ & $0.4150$ \\ 
\hline \hline
\end{tabular}
\end{table}

\subsection{Evolution of Priority Patterns via Quadrant Analysis}

Here we investigate the relationship between the excess degree of Section~\ref{sec:significant} and a paper's citation impact. 
We categorize the nodes into four groups based on the signs of their excess degrees: Q1 ($S(H) > 0$ \& $S(I) > 0$), Q2 ($S(H) < 0$ \& $S(I) > 0$), Q3 ($S(H) < 0$ \& $S(I) < 0$), and Q4 ($S(H) > 0$ \& $S(I) < 0$). 
Subsequently, we measure how the proportions of these groups evolve as a function of the in-degree threshold $k_{in}$.

\begin{figure}[ht]
\centering
\includegraphics[angle=0,width=1\columnwidth]{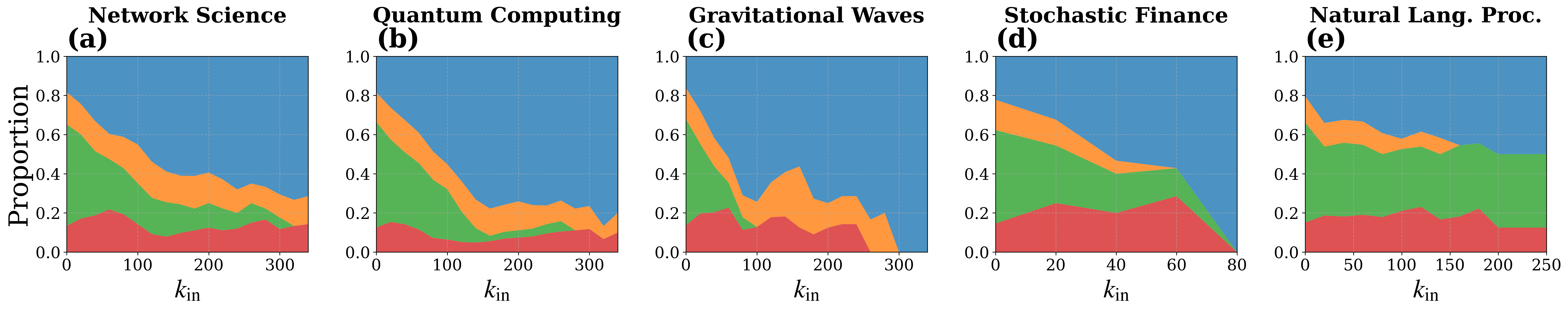}
\caption{Relation between the consistency of reference selection for the two different prompts and the number of citations of the references. Papers are divided into four groups, according to the signs of their excess degrees. Each panel corresponds to one of the five fields we have considered in the analysis:
(a) Network Science, (b) Quantum Computing, (c) Gravitational Waves, (d) Stochastic Finance, and (e) Natural Lang. Proc.. The stacked area plots illustrate the variable relative shares of each group as the network is filtered for nodes with progressively higher in-degrees. The colors designate specific groups: Q1 (blue), Q2 (orange), Q3 (green), and Q4 (red).
}
\label{fig:quadrant}
\end{figure}

Figure~\ref{fig:quadrant}(a-c) illustrates that two fields (Network Science and Quantum Computing) exhibit qualitatively similar results.
As the threshold $k_{in}$ increases, the proportion of Q1 expands, while Q3 gradually decreases and eventually vanishes. Notably, the proportions of Q2 and Q4 remain stable even at high $k_{in}$. Gravitational Waves (Fig.~\ref{fig:quadrant}(c)) has a similar pattern, but only Q1 survives when $k_{in} > 200$. 
Therefore, while highly cited papers are often selected by the LLM with the two different prompts, the LLM does not systematically prioritize all high-impact works; rather, it evaluates the specific qualitative role of each citation, often selecting a hub for one prompt while excluding it from the other based on its contextual relevance.

In the other fields, different tendencies are observed. For Stochastic Finance (Fig.~\ref{fig:quadrant}(d)), when $k_{in} > 80$, the remaining papers are eventually located in Q1. In contrast to the results above, Q2 is the first to disappear, while Q3 and Q4 persist even at high citation thresholds.
A similar phenomenon is observed for Natural Language Processing (Fig.~\ref{fig:quadrant}(e)), where Q2 shrinks first as $k_{in}$ increases, while Q3 and Q4 keep their proportions.
These results imply that even if a paper has received numerous citations, it is not necessarily selected by the LLM if its citation context does not align with the specific intent of the prompt.

\subsection{Structural Analysis of Base References}

The previous analyses have established that the LLM's reference selection is neither random nor too biased toward highly cited papers.
To investigate how these selections are reflected by the position of the references in the body of the paper, we analyze the distribution of the base references across the paper's sections.
Detailed information on the section extraction methodology and the criteria for structural classification is provided in Section~\ref{m:section}.

\begin{figure}[ht]
\centering
\includegraphics[angle=0,width=1\columnwidth]{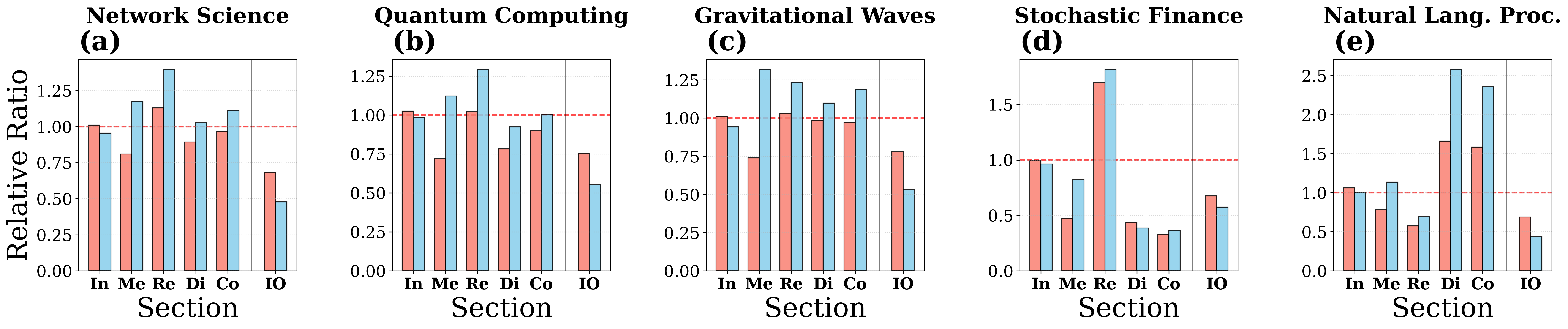}
\caption{
Relative ratio of references across paper sections. The bars indicate the relative ratio of citation fractions in the backbone networks $H$ (red) and $I$ (blue) compared to the full-citation network ($G$).
For the primary sections (In, Me, Re, Di, Co), each proportion is calculated relative to the sum of citations across all five sections. In contrast, the IO category represents an internal ratio, defined specifically as the number of references cited exclusively in the Introduction divided by the total number of references within the Introduction section.
Values above (below) $1.0$ indicate that the LLM selects references from those sections more (less) frequently with respect to their overall frequency in the field.
Panels represent five scientific domains: (a) Network Science, (b) Quantum Computing, (c) Gravitational Waves, (d) Stochastic Finance, and (e) Natural Language Processing.
The details of section extraction are given in Sec.~\ref{m:section}.
}
\label{fig:section}
\end{figure}

Figure~\ref{fig:section} shows the structural distribution of base references across all five fields. For each section, we compute the ratio between the fraction of references in a certain section in $H$ or $I$ and the corresponding fraction in the full citation network $G$.
Notably, across all fields, the relative ratio for the \textit{Introduction} section is consistently slightly higher in $H$ compared to $I$. This indicates that the ``Significant references'' a bit more concentrated in the \textit{Introduction} section of the papers than those selected as ``Inspirational References''.
In four fields (except for Stochastic Finance), the references of papers in $I$ appear more frequently in the \textit{Methods}, \textit{Results}, \textit{Discussion}, and \textit{Conclusion} compared to those of $H$. For the backbone network $I$, these relative ratios frequently exceed $1.0$, indicating that the LLM prioritizes ``Inspirational references'' not only as background but also for their role in the core technical sections of the paper.
A distinct pattern is observed for Stochastic Finance. 
While the relative ratio of $H$ in the \textit{Discussion} is larger than that of $I$, the relative ratios of $I$ remain more pronounced in the \textit{Methods}, \textit{Results}, and \textit{Conclusions}, consistently with the other fields.

We also observe several field-specific outliers that highlight diverging citation cultures. In Stochastic Finance, the relative ratio for the \textit{Results} is exceptionally high ($>1.5$) in both backbone networks, whereas the ratios for the \textit{Discussion} and \textit{Conclusions} are notably suppressed ($<0.5$). The qualitatively important references prioritized by both criteria (``Significant references'' and ``Inspirational references'') are concentrated within the analytical and empirical outcomes of the study.
In contrast, Natural Language Processing exhibits a shift toward the final stages of the paper; the relative ratios for the \textit{Discussion} and \textit{Conclusions} are prominent, exceeding $1.5$ for $H$ and $2.0$ for $I$. This indicates that in this field, the base references prioritized by both criteria are primarily situated in the interpretation of results and the contextualization of findings within the broader research landscape.

Furthermore, to gain deeper insight into the LLM’s selection logic, we conducted an additional analysis of the \textit{Introduction-Only} category.
As shown in Fig.~\ref{fig:section}, the relative ratios are consistently below $1.0$ across all fields, with $H$ maintaining a higher ratio than $I$. This suggests that the LLM tends to de-prioritize references that serve exclusively as background information within the introduction. Instead, the LLM prioritizes references that exhibit multi-sectional presence, specifically those that are introduced in the \textit{Introduction} and subsequently utilized in other sections of the paper.



Based on these results, the relative ratios of references prioritized by both prompts confirm that each prompt selects references occupying a characteristic structural position, reflecting how those references are used across sections. 
The prompt designed to identify ``significance'' leads to a more frequent selection of references from the \textit{Introduction}.
In contrast, the prompt designed to identify ``inspirational references'' focuses on the foundational methodology or core ideas, which results in the higher prevalence of references in the \textit{Methods}, \textit{Results}, \textit{Discussion}, and \textit{Conclusions} sections. The differences are not big but sizable.

Beyond these prompt-driven differences, our analysis also captures distinct disciplinary nuances; notably, the `results-centric' pattern in Stochastic Finance and the `conclusion-heavy' distribution in Natural Language Processing highlight how the focus shifts across fields according to their citation cultures.
These results demonstrate that the LLM-based backbone extraction effectively distills the qualitative `intellectual anchors' of a field, grounded in the specific structural narrative of the scientific articles.


\subsection{Analysis of Author Citation Networks}

\begin{table}[ht]
\centering
\caption{Summary statistics of author citation networks ($G_a$, $H_a$, and $I_a$) corresponding to the full paper citation network and backbone networks.}
\label{tab:author_net_info}
\begin{tabular}{lllll}
\hline \hline
   & $N$ & $E(G_a)$ & $E(H_a)$ & $E(I_a)$ \\ \hline
Network Science      & $22\,154$  &  $355\,834$    &   $86\,774$   &  $81\,755$    \\
Quantum Computing    & $21\,435$  &  $702\,753$    &   $148\,584$   &  $150\,630$    \\
Gravitational Waves  & $33\,530$  &  $506\,844$    &   $100\,897$   &  $94\,047$    \\
Stochastic Finance   & $9\,358$  &  $33\,585$   &   $8\,987$   &  $10\,030$ \\
Natural Lang. Proc.  & $12\,936$  &  $332\,712$   &   $79\,450$   &  $81\,301$    \\
\hline \hline
\end{tabular}
\end{table}

\begin{table}[ht]
\centering
\caption{
Pearson ($\rho_{\mathrm{P}}$) and Spearman ($\rho_{\mathrm{S}}$) correlation coefficients of in-strength (total citations) in author citation networks ($G_a$, $H_a$, and $I_a$).
Overall, the correlation coefficients exhibit consistently high values across all networks.
The correlation between backbone networks $H_a$ and $I_a$ is remarkably strong.
}
\label{tab:correlation_instrength}
\begin{tabular}{llllllll}
\hline \hline
    & $\rho_{\mathrm{P}}(G_a, H_a)$ & $\rho_{\mathrm{P}}(G_a, I_a)$ & $\rho_{\mathrm{P}}(H_a, I_a)$ & $\rho_{\mathrm{S}}(G_a, H_a)$ & $\rho_{\mathrm{S}}(G_a, I_a)$ & $\rho_{\mathrm{S}}(H_a, I_a)$ \\ \hline
Network Science & $0.9720$ & $0.9781$ & $0.9899$ & $0.7288$ & $0.7275$ & $0.7695$\\
Quantum Computing & $0.9208$ & $0.9332$ & $0.9211$ & $0.7737$ & $0.7695$ & $0.7847$\\
Gravitational Waves & $0.9359$ & $0.9478$ & $0.9477$ & $0.7156$ & $0.7082$ & $0.7333$\\
Stochastic Finance & $0.9528$ & $0.9495$ & $0.9646$ & $0.7286$ & $0.7629$ & $0.7779$\\
Natural Lang. Proc. & $0.9600$ & $0.9509$ & $0.9524$ & $0.7719$ & $0.7645$ & $0.7928$\\
\hline \hline
\end{tabular}
\end{table}

\begin{table}[ht]
\centering
\caption{
Pearson ($\rho_{\mathrm{P}}$) and Spearman ($\rho_{\mathrm{S}}$) correlation coefficients of PageRank in author citation networks ($G_a$, $H_a$, and $I_a$).
Overall, the correlation coefficients exhibit consistently high values across all networks.
The correlation between backbone networks $H_a$ and $I_a$ is remarkably strong.
}
\label{tab:correlation_PageRank}
\begin{tabular}{llllllll}
\hline \hline
    & $\rho_{\mathrm{P}}(G_a, H_a)$ & $\rho_{\mathrm{P}}(G_a, I_a)$ & $\rho_{\mathrm{P}}(H_a, I_a)$ & $\rho_{\mathrm{S}}(G_a, H_a)$ & $\rho_{\mathrm{S}}(G_a, I_a)$ & $\rho_{\mathrm{S}}(H_a, I_a)$ \\ \hline
Network Science & $0.9703$ & $0.9578$ & $0.9799$ & $0.7188$ & $0.7212$ & $0.7656$\\
Quantum Computing & $0.8178$ & $0.8163$ & $0.9144$ & $0.7630$ & $0.7601$ & $0.7788$\\
Gravitational Waves & $0.8830$ & $0.9003$ & $0.9048$ & $0.7630$ & $0.7601$ & $0.7788$\\
Stochastic Finance & $0.8041$ & $0.7410$ & $0.7893$ & $0.7204$ & $0.7551$ & $0.7734$\\
Natural Lang. Proc. & $0.9237$ & $0.9336$ & $0.9895$ & $0.7586$ & $0.7574$ & $0.7893$\\
\hline \hline
\end{tabular}
\end{table}

\begin{table}[ht]
\centering
\caption{Comparison of the top 10 scientists in Network Science, ranked by in-degree in $G_a$, $H_a$, and $I_a$. The table comprises the union of the top 10 lists from each network and is sorted by the in-degree in $G_a$, $k(G_a)$.}
\label{tab:rank_scientist}
\begin{threeparttable}
\begin{tabularx}{0.7\textwidth}{l ccc}
\hline \hline

  Name  & $k(G)$ (Rank) & $k(H)$ (Rank) & $k(I)$ (Rank)  \\ \hline
  M. E. J. Newman & $3\,917$ (1) &$946$ (1) & $959$ (1) \\
  Shlomo Havlin & $3\,040$ (2) & $863$ (3) & $717$ (4) \\
  Albert Laszlo Barabasi & $2\,664$ (3) & $937$ (2) & $843$ (2) \\
  Réka Albert & $2\,630$ (4) &  $780$ (4) & $786$ (3) \\
  Romualdo Pastor-Satorras & $2\,565$ (5) & $533$ (5) & $472$ (5) \\
  Alessandro Vespignani & $2\,044$ (6) & $497$ (6) & $410$ (6) \\
  S. N. Dorogovtsev  & $1\,747$ (7) & $400$ (10) & $372$ (8) \\
  José F. F. Mendes & $1\,713$ (8) & $396$ (10+) & $368$ (9) \\
  Santo Fortunato & $1\,554$ (9) & $407$ (9) & $324$ (10+) \\
  H. Eugene Stanley & $1\,481$ (10) & $465$ (7) & $399$ (7) \\
  Sergey V. Buldyrev & $1\,249$ (10+) & $423$ (8) & $341$ (10) \\

\hline \hline
\end{tabularx}
\end{threeparttable}
\end{table}

To extend our analysis from paper-level structures to researcher interactions, we construct author-level citation networks for the full citation and backbone networks. Utilizing the unique OpenAlex author identifiers available for all considered papers, we establish a mapping between papers and their respective authors. In these networks, nodes represent individual authors, and directed edges denote citation relationships. Specifically, a directed edge goes from author $A$ to author $B$ if $A$ cites at least one paper written by $B$. To capture the volume of citations, these edges are weighted; for instance, if author $A$ cites three distinct papers by author $B$, the directed edge weight is three. We exclude all self-citations. An author's total citation count corresponds to their node in-strength, defined as the sum of the weights of their incoming edges. Table~\ref{tab:author_net_info} summarizes the number of nodes and edges in the three author citation networks $G_a$, $H_a$, and $I_a$. As expected, the full-citation networks contain significantly more edges than the backbone networks.

Similarly to the results shown in Tabs.~\ref{tab:Pearson} and~\ref{tab:Spearman}, we quantify the correlation of the nodes' in-strength~\cite{REF_Pearson,REF_Spearman} in the full network and the backbones. According to Tab.~\ref{tab:correlation_instrength}, the correlations are remarkably strong. Furthermore, we evaluate the correlations of the authors' PageRank values~\cite{PageRank}. Unlike direct in-strength counts, PageRank employs Markov chains to incorporate information from extended neighborhoods. However, as shown in Tab.~\ref{tab:correlation_PageRank}, the differences between the full citation and backbone networks remain minimal even under this metric. These strong correlations demonstrate that the consistency between the full citation network and the backbones previously found at the paper level is maintained at the author level.

\section{Discussion}

In this work we have focused on the ``important" references of papers. The goal was twofold. On the one hand, we wanted to highlight the sources of a paper, namely the references that led to the design and development of the research of the paper. On the other hand, we wanted to eliminate redundancy, as much as possible, and estimate credit to papers and authors based on the actual sources. We used an LLM to extract the sources from the list of references of papers in five different scientific fields. The resulting citation networks could be considered backbones of the initial (full) citation networks.
We used two different prompts for the LLM, asking the model to single out ``significant'' and ``inspirational'' references, respectively.

We made several interesting observations. First, while we did not impose constraints on the number of selected references, the LLM usually extracted just a few references per paper, and left most of the original ones out. Hence, backbone networks end up being much sparser than the corresponding full networks. Second, the in-degree distributions of the backbones are qualitatively similar to the ones of the full network, in that they are very skewed and with approximately power-law tails, with the backbones displaying slightly smaller exponents and, consequently, higher heterogeneity. Also, the correlation between in-degrees of papers in the full citation network and the backbones is very high, signaling that rankings of papers by their number of citations mostly reflect their actual citation impact when only the sources are considered. Third, there are differences in other structural features such as the the profiles of the clustering coefficient and the degree of nearest-neighbors versus the degree (in the undirected versions of the networks). In particular, backbone networks are much less modular than the corresponding full networks, which suggests that edges to the sources are more evenly placed in the network than edges to the other references, which may tend to be more concentrated within subtopics. Fourth, we introduced a variable, that we called excess degree, to estimate how much more (or less) cited a paper is in the backbone with respect to an equivalent random parsing of the initial full network, and found that the distribution of excess degree, while concentrated around zero, stretches towards large positive and negative values, indicating that there are papers which fare much better or much worse than they would if references were randomly removed. This suggests that the LLM is not simply picking references randomly, a result that we have also directly verified by extending the structural analysis of the networks to randomly pruned versions of the backbones with equal sparsity. Fifth, the two prompts we used lead to very similar networks from a structural viewpoint, but with significant differences when it comes to the selection of the references. In particular, there are papers that are frequently selected using one prompt, and not so frequently using the other. A notable example is the paper by Blondel et al. that introduced the Louvain algorithm for community detection in network science~\cite{blondel08}. This paper, among the most cited in the field, is much less prominent when the LLM is asked to select ``inspirational'' references. The reason is that the algorithm itself did not inspire much research, but it is frequently used as a core methodological tool in applications. Indeed, we found that ``significant'' references are more often drawn from the Introduction section, whereas the ``inspirational'' ones are more often drawn from the technical sections of the paper.
Lastly, when one considers citations between authors, rankings of authors according to both the raw number of citations and PageRank are very consistent between the full network and the backbone(s), confirming that constraining citations to papers and their sources does not alter the perception of relative importance of authors that one derives from the complete system.

An important caveat for this work concerns the use of LLMs to extract the critical references from a paper's bibliography. While it is currently the only way to execute this task at scale, and LLMs have proven to be generally very reliable when it comes to assessing context, there is no guarantee that sources can be extracted this way. 
This is why we have used two slightly distinct prompts, which, despite selecting not highly overlapping sets of references, delivered very similar backbones from the structural point of view. Initial manual tests we have carried out initially on few dozens of papers indicated that the procedure, while imperfect, is still reliable. 
A more robust analysis would require using multiple prompts and different LLMs (we only used one), but it would take a lot of time and computational resources.

\section{Methods}

\subsection{Data}\label{m:data}

\textbf{Data Sources and Primary Dataset.}\, To obtain the full text of scientific articles, we used the unarXive dataset~\cite{unarxive}. This dataset provides a comprehensive collection of arXiv pre-prints, including preprocessed reference strings and their corresponding OpenAlex identifiers (IDs), based on the OpenAlex snapshot downloaded in December 2024. Using unarXive, we can access structured textual data along with the initial citation metadata. To maintain compatibility with the context window constraints of the LLM, the dataset is further refined by excluding any articles with a full-text token count exceeding $100\,000$ (about 50 pages) or those lacking a reference list.

\textbf{Bibliographic Enrichment and Reference Matching.}\, While unarXive provides OpenAlex IDs for many references, the metadata is occasionally incomplete, with some entries lacking titles or unique identifiers. To address these gaps and enhance the citation network's coverage, we employ AnyStyle, a Ruby-based bibliographic parser~\cite{anystyle}, to extract granular metadata such as titles, volumes, and page numbers. We query the Crossref API~\cite{crossref} using these parsed strings to retrieve missing paper titles. To establish an edge (citation) between two papers and reduce the false positives, we perform a strict matching process: a citation is confirmed if both the title and publication year of a reference precisely matched those of a pre-print in the dataset.

\textbf{Field Selection and Network Construction.}\, 
The study identifies five research fields using their respective OpenAlex Topic IDs. The Network Science field incorporates three distinct topics: Complex Network Analysis Techniques (T10064), Nonlinear Dynamics and Pattern Formation (T11187), and Opinion Dynamics and Social Influence (T12592). Other analyzed fields include Quantum Computing Algorithms and Architecture (Quantum Computing; T10682), Pulsars and Gravitational Waves Research (Gravitational Waves; T10463), Natural Language Processing Techniques (Natural Lang. Proc.; T10181), and Stochastic Processes and Financial Applications (Stochastic Finance; T10067).
We chose these five fields because they represent very different areas of research.
A significant challenge in citation analysis is the existence of multiple versions of the same work (e.g., an arXiv pre-print and a subsequent peer-reviewed journal publication). Since researchers cite the published version, we implement an entity resolution step to map OpenAlex IDs of published papers to their corresponding pre-prints in the unarXive dataset. This ensures that the resulting directed networks accurately reflect the citation flow, regardless of which version was formally cited.

\subsection{LLM-based Information Extraction}\label{m:LLM}

\textbf{Model Configuration.}\,
For the reference extraction task, we adapt the \textit{DeepSeek-R1-Distill-Llama-70B} model, implemented on the vLLM serving platform to ensure high-throughput inference~\cite{deepseek,vllm}. 
Given that our objective is precise information extraction rather than creative generation, we configured the model with a low temperature and top-p (both set to $0.1$).
These deterministic settings are chosen to minimize stochasticity and maximize the reproducibility of the extracted citation sets.

Furthermore, supplementary tests using a more compact variant, \textit{DeepSeek-R1-Distill-Llama-8B}, produce qualitatively consistent results in the Network Science field. This consistency suggests that the observed patterns remain robust across different model scales.

\textbf{Input representation and bias mitigation.}\,
To ensure that the extraction process strictly relies on the functional role of citations within the provided text, we represent all references solely by their numeric markers (e.g., [1], [15]) as they appear in the full-text. Crucially, the full texts omit all bibliographic metadata, such as paper titles, author names, or DOI. This anonymization strategy is designed to mitigate potential biases arising from the LLM’s pre-training data; by withholding the identity of the cited works, we force the model to evaluate a reference based on its contextual necessity and the author's reasoning within the main text, rather than the established reputation or citation impact of the paper.

\textbf{Two-Step Extraction and Parsing Pipeline.}\,
To ensure the integrity of the data, we decouple the reasoning-based extraction from the final data structuring process through a systematic two-step pipeline:
\begin{enumerate}[label=\textbf{\arabic*.}, leftmargin=2cm, labelsep=0.5cm]
    \item \textbf{Reasoning and Initial Extraction:} In the first stage, the LLM identifies references that met the criteria as detailed in \textbf{Prompts} and Appendices~\ref{prompt1} and ~\ref{prompt2}. The LLM provides both the reference numbers and the underlying rationale for each selection. At the end of its response, it provides a consolidated summary list of the identified references in plain text format.
    \item \textbf{Structured Post-Processing:} To translate the naively structured list, we implement a subsequent parsing step. This step utilizes the LLM to take the consolidated list from the first response and generate a structured output consisting solely of the validated reference numbers.
\end{enumerate}
This two-stage approach isolates qualitative reasoning from the final formatting step. 

\textbf{Prompts.}\,
The extraction process described in this Section operates based on two distinct analytical dimensions to identify the references for the backbone networks $H$ and $I$, respectively.
These dimensions define the specific ``\{category\_name\}'' and ``\{category\_description\}'' provided to the LLM to guide the filtering of citations. The exact inputs for each dimension are as follows, where the term in bold represents the category name and the quoted text represents its description:
\begin{enumerate}[label=\textbf{\arabic*.}, leftmargin=2cm, labelsep=0.5cm]
    \item[$H$.] \textbf{Significant references:} Identify cited references that help to clearly establish the broader background of the research problem this paper addresses, the existing landscape or current state of the field, or the primary research gap it aims to fill. This category also includes references cited to underscore why this research topic is important and to highlight the necessity for this study. Crucially, focus on references that provide the essential context for readers to understand the paper's research motivations and its potential academic and/or social contributions.
    \item[$I$.] \textbf{Inspirational references:} Identify and list the references that inspired this article.
\end{enumerate}
The inference and parsing stages utilize these definitions within a structured prompt framework. Notably, the prompt intentionally omits any requirement regarding the number of references to be extracted. By refraining from setting a fixed quota, the methodology ensures that the resulting backbone networks reflect the intrinsic relevance of the citations rather than an arbitrary numerical constraint. To ensure transparency, Appendices~\ref{prompt1} and~\ref{prompt2} provide the full textual templates of the prompts, including the exact category labels and descriptions used during extraction.

\subsection{Network Definitions and Baselines}\label{m:network}

\textbf{Definition and Boundary.}\,
The constructed networks are directed, since they are based on citations. To maintain structural consistency, we excluded all external citations, restricting the edge set exclusively to references between papers within each field.

\textbf{Edge Directionality for Metric Calculation.}\,
To measure the degree-dependent clustering coefficient, average normalized degree of nearest-neighbors, and robustness modularity, we consider edges as undirected.

\textbf{Baselines of Backbone Networks.}\,
The generation of baseline networks ($H_r$, $H_m$, etc.) relies on the structural properties of the original backbone networks.
Each baseline preserves the exact out-degree (the number of extracted references) of every paper while altering the selection logic:
\begin{enumerate}[label=\textbf{\arabic*.}, leftmargin=2cm, labelsep=0.5cm]
    \item \textbf{Random baselines ($H_r, I_r$):} For each individual paper, references are stochastically chosen from its original reference list in the full citation network $G$. Crucially, the number of selected references per paper is strictly constrained to match the exact number of references retained in the original backbone network. 
    \item \textbf{Most-cited baselines ($H_m, I_m$):} For each individual paper, the most cited references are selected from its original reference list in the full citation network $G$. As with the random baseline, the number of selected references per paper is strictly constrained to match the exact number of references retained in the original backbone network. 
\end{enumerate}
We stress that the baselines are in principle the same for both $H$ and $I$, as they are obtained by operating on the full reference lists of the papers. However, for any given paper each prompt returns a different number of references for $H$ and $I$, in general, and we impose that the baselines have the mean in-degree (out-degree) as $H$ and $I$, hence the two different pairs of networks ($H_r, I_r$) and ($H_m, I_m$).

\subsection{Classification of Structural Sections}\label{m:section}

To analyze the structural context of references, we focus on a subset of preprints that contain explicit section metadata. Given the high variability in section naming conventions across diverse scientific literatures, the analysis employs a keyword-based normalization strategy to categorize sections into five primary types: \textit{Introduction} (In), \textit{Methods} (Me), \textit{Results} (Re), \textit{Discussion} (Di), and \textit{Conclusions} (Co).

Under this classification scheme, a section is assigned to a category if its title contains the keyword; a section titled ``Method and Theory'' is grouped under the Methods category. This approach ensures a consistent framework for comparison across different paper structures.

As a single reference may be cited across multiple parts of an article, the aggregate count of occurrences across sections very often exceeds the total number of unique papers. This is particularly frequent for references that are first mentioned in the introduction and subsequently cited in technical or analytical sections.
To distinguish between such cases and those providing only general background information, the category \textit{Introduction-Only} (IO), is defined for references cited exclusively within the introductory section.

To quantify the structural prioritization of the LLM as presented in Fig.~\ref{fig:section}, we calculate a relative ratio for these categories. We first determine the proportion $P$ of references for each section in the full network ($G$) and the backbone networks ($H$ and $I$). For the primary sections ($i \in \{ \mathrm{In, Me, Re, Di, Co} \}$), the proportion is calculated relative to their combined total:
\begin{equation}
P_{i}(x) = \frac{\nu_{i}(x)}{\sum_{j \in \{ \mathrm{In, Me, Re, Di, Co}\}} \nu_{j}(x)} ,
\end{equation}
where $\nu_{i}(x)$ denotes the number of occurrences of references in a given section for network $x \in \{G, H, I\}$.

In contrast, to analyze the internal composition of the \textit{Introduction}, the \textit{Introduction-Only} category is defined as the proportion of references in the \textit{Introduction} that appear only there:
\begin{equation}
P_{IO}(x) = \frac{\nu_{\mathrm{IO}(x)}}{\nu_{\mathrm{In}}(x)} .
\end{equation}

Finally, the relative ratio is computed by dividing the backbone proportions by the corresponding baseline in $G$:
\begin{equation}
\text{Relative Ratio} = \frac{P(H)}{P(G)} \quad \text{or} \quad \frac{P(I)}{P(G)} .
\end{equation}
A ratio greater than $1.0$ indicates that the LLM-selected references are over-represented in that specific category compared to the field's average proportion.

\bibliographystyle{plain}
\bibliography{References}

\section*{Acknowledgments}
D.M. and S.F. acknowledge the support of the AccelNet-MultiNet program, a project of the National Science Foundation (Awards \#1927425 and \#1927418). S. F. and W. J. acknowledge the support of the grant NNF24SA0092140 from the Novo Nordisk Foundation. D. M. also acknowledges the support from the Spanish grants PID2021-128005NB-C22 and PID2024-158120NB-C22, funded by MCIN/AEI/10.13039/501100011033 and “ERDF A way of making Europe".

\appendix
\section*{Appendix}

\section{Prompt for the Inference}\label{prompt1}

Below we report the prompts for the initial parsing of the paper content. See Methods, Section~\ref{m:LLM}

\begin{tcblisting}{
    colback=white, 
    colframe=black, 
    arc=0mm, 
    title=Inference,
    listing only,
    breakable,        % over the page
    enhanced,         % for the breakable
    listing options={
        basicstyle=\small\ttfamily, 
        breaklines=true,   % break line
        columns=flexible, 
        keepspaces=true
    }
}
You are an expert in scientific article analysis. Your current task is to perform an in-depth textual analysis of the "Given article" focusing *only* on the category: "{category_name}".

### Given article:
{article_text}

### References:
All references are cited by the number (e.g, [1],[3],[12])


### Instructions for Analysis (Part 1 - Textual Output Only):

1.  **Focus Area Definition:**
    Thoroughly analyze the "Given article". Your analysis must concentrate *exclusively* on aspects relevant to the category "{category_name}".
    The definition for "{category_name}" is: "{category_description}".

2.  **Selected Reference Identifiers (Refs):**
    After detailing your CoT, clearly list the reference numbers (e.g., "1", "3", "12") you have selected as most critical and relevant for "{category_name}". These should be only the reference numbers/identifiers.

3.  **Detailed Reasons for Each Selected Reference (Reasons):**
    For *each* reference number you listed in the step above, provide a detailed and specific explanation (e.g., 1-2 complete sentences per reference) stating *precisely why* that particular reference is critically important or highly relevant to "{category_name}" according to your analysis and the category definition. Each reason must correspond to a selected reference.

### Required Output Format for This Step (Text Only):

Please structure your entire response for this step clearly under the following three distinct headers. Do not include any other text outside these sections.

`SELECTED_REFERENCES_FOR_{header_category_name}:`
[List only the selected reference numbers here, one per line, or comma-separated. For example:
"1"
"3"
"12"
or "1", "3", "12"]

`DETAILED_REASONS_FOR_{header_category_name}:`
[Provide one detailed reason for each selected reference, ensuring the order matches the SELECTED_REFERENCES_FOR_{header_category_name} list. Each reason should start on a new line or be clearly delineated. For example:
Reason for reference "1": [Detailed reason specific to {category_name}]
Reason for reference "3": [Detailed reason specific to {category_name}]
Reason for reference "12": [Detailed reason specific to {category_name}]]
\end{tcblisting}

\clearpage

\section{Prompt for the Parsing}\label{prompt2}

Below we report the prompts to obtain the structured output for the output of the first prompt. See Methods, Section~\ref{m:LLM}

\begin{tcblisting}{
    colback=white, 
    colframe=black, 
    arc=0mm, 
    title=Parsing,
    listing only,
    breakable,        % over the page
    enhanced,         % for the breakable
    listing options={
        basicstyle=\small\ttfamily, 
        breaklines=true,   % break line
        columns=flexible, 
        keepspaces=true
    }
}
You are an expert utility that converts structured textual analysis into a specific JSON format.
You will be provided with a block of text generated by a previous analytical step. This text contains three key sections:
1.  A list of selected reference identifiers.
2.  A list of detailed reasons explaining the relevance of each selected reference to that category.

Your task is to parse this input text and accurately convert it into a single JSON object with the keys "refs", and "reasons".

### Input Text (contains selected references and detailed reasons):
{output_text_of_first_trial}

### Instructions for JSON Generation:

1.  **Extract "refs" List:**
    Identify the list of reference numbers under a header similar to "SELECTED_REFERENCES_FOR_...".
    Parse these numbers and format them as a list of strings in the JSON "refs" field. For example, if the text lists "1", "15", "42", the JSON field should be `["1", "15", "42"]`. Ensure only the reference numbers/identifiers are included.

2.  **Extract "reasons" List:**
    Identify the detailed reasons provided under a header similar to "DETAILED_REASONS_FOR_...".
    Each reason corresponds to a reference in the "refs" list. Parse these reasons and format them as a list of strings for the JSON "reasons" field.
    The order and number of reasons must strictly match the order and number of items in the "refs" list. Each reason should be a complete thought explaining the relevance.

### Output must **only** be in the following JSON format:
{{
    "refs": ["<ref_id_1_str>", "<ref_id_2_str>", ...],
    "reasons": ["<detailed_reason_for_ref_1>", "<detailed_reason_for_ref_2>", ...]
}}

Ensure your output is a single, valid JSON object and nothing else. Do not add any explanatory text before or after the JSON.
\end{tcblisting}

\clearpage

\section{Rank of papers}\label{appendix_rank}

Below we report the tables with the rank of the papers in different disciplines.

\begin{table}[ht]
\centering
\caption{Comparison of the top 10 papers in the Quantum Computing, ranked by in-degree in $G$, $H$, and $I$. The table comprises the union of the top 10 lists from each network and is sorted by the in-degree in $G$, $k(G)$.}
\label{tab:rank2}
\small
\begin{tabularx}{\textwidth}{X c ccc}
\hline \hline

  Title\tnote{a}  & Year & $k(G)$ (Rank) & $k(H)$ (Rank) & $k(I)$ (Rank)  \\ \hline
  Quantum Computing in the NISQ era and beyond & 2018 & $1\,225$ (1) &$712$ (1) & $225$ (2) \\
  A Quantum Approximate Optimization Algorithm & 2014 & $834$ (2) & $312$ (2) & $364$ (1) \\
  Elementary gates for quantum computation & 1995 & $660$ (3) & $131$ (10) & $147$ (5) \\
  Stabilizer Codes and Quantum Error Correction & 1997 & $583$ (4) & $188$ (4) & $194$ (3) \\
  Good Quantum Error-Correcting Codes Exist & 1995 & $449$ (5) & $141$ (5) & $154$ (4) \\
  Supplementary information for "Quantum supremacy using a programmable superconducting processor" & 2019 & $389$ (6) & $204$ (3) & $97$ (10+) \\
  A variational eigenvalue solver on a quantum processor & 2013 & $375$ (7) & $134$ (8) & $146$ (6) \\
  Quantum algorithm for solving linear systems of equations & 2008 & $344$ (8) & $135$ (7) & $126$ (7) \\
  A Theory of Quantum Error-Correcting Codes & 1996 & $343$ (9) & $112$ (10+) & $117$ (10) \\
  Quantum random walks - an introductory overview & 2003 & $342$ (10) & $110$ (10+) & $92$ (10+) \\
  Quantum Computation and Decision Trees & 1997 & $333$ (10+) & $105$ (10+) & $119$ (8) \\
  Quantum Machine Learning & 2016 & $323$ (10+) & $135$ (6) & $76$ (10+) \\
  Barren plateaus in quantum neural network training landscapes & 2018 & $326$ (10+) & $131$ (9) & $89$ (10+) \\
  Fault-tolerant quantum computation by anyons & 1997 & $260$ (10+) & $120$ (10+) & $119$ (8) \\

\hline \hline
\end{tabularx}
\end{table}

\begin{table}[ht]
\centering
\caption{Comparison of the top 10 papers in the Gravitational Waves, ranked by in-degree in $G$, $H$, and $I$. The table comprises the union of the top 10 lists from each network and is sorted by the in-degree in $G$, $k(G)$.}
\label{tab:rank3}
\small
\begin{tabularx}{\textwidth}{X c ccc}
\hline \hline

  Title  & Year & $k(G)$ (Rank) & $k(H)$ (Rank) & $k(I)$ (Rank)  \\ \hline
  Evolution of Binary Black Hole Spacetimes & 2005 & $374$ (1) &$197$ (1) & $172$ (1) \\
  Accurate Evolutions of Orbiting Black-Hole Binaries Without Excision & 2005 & $324$ (2) & $108$ (4) & $121$ (2) \\
  Tests of general relativity with GW150914 & 2016 & $302$ (3) & $146$ (2) & $59$ (8) \\
  The equation of state for nucleon matter and neutron star structure & 1998 & $295$ (4) & $45$ (10+) & $68$ (6) \\
  Gravitational wave extraction from an inspiraling configuration of merging black holes & 2005 & $291$ (5) & $113$ (3) & $111$ (4) \\
  Effective one-body approach to general relativistic two-body dynamics & 1998 & $274$ (6) & $96$ (5) & $121$ (3) \\
  GW170608: Observation of a 19-solar-mass Binary Black Hole Coalescence & 2017 & $243$ (7) & $56$ (8) & $26$ (10+) \\
  Transition from inspiral to plunge in binary black hole coalescences & 2000 & $202$ (8) & $51$ (10+) & $78$ (5) \\
  The Physics of Neutron Stars & 2004 & $191$ (9) & $70$ (7) & $39$ (10+) \\
  Constraining neutron star tidal Love numbers with gravitational wave detectors & 2007 & $187$ (10) & $75$ (6) & $65$ (7) \\
  Coalescence of Two Spinning Black Holes: An Effective One-Body Approach & 2001 & $155$ (10+) & $41$ (10+) & $56$ (9) \\
  Is the gravitational-wave ringdown a probe of the event horizon? & 2016 & $134$ (10+) & $55$ (9) & $43$ (10+) \\
  Pulsars as Astrophysical Laboratories for Nuclear and Particle Physics & 2006 & $119$ (10+) & $54$ (10) & $29$ (10+) \\

  Gravitational Radiation Reaction to a Particle Motion & 1996 & $103$ (10+) & $29$ (10+) & $49$ (10) \\
  
\hline \hline
\end{tabularx}
\end{table}

\begin{table}[ht]
\centering
\caption{Comparison of the top 10 papers in the Stochastic Finance, ranked by in-degree in $G$, $H$, and $I$. The table comprises the union of the top 10 lists from each network and is sorted by the in-degree in $G$, $k(G)$.}
\label{tab:rank4}
\small
\begin{tabularx}{\textwidth}{X c ccc}
\hline \hline

  Title  & Year & $k(G)$ (Rank) & $k(H)$ (Rank) & $k(I)$ (Rank)  \\ \hline
  The master equation and the convergence problem in mean field games & 2015 & $89$ (1) & $30$ (2) & $34$ (2) \\
  Well-posedness of the transport equation by stochastic perturbation & 2008 & $85$ (2) & $33$ (1) & $36$ (1) \\
  Strong convergence of an explicit numerical method for SDEs with nonglobally Lipschitz continuous coefficients & 2010 & $76$ (3) & $25$ (4) & $28$ (4) \\
  A note on tamed Euler approximations & 2013 & $67$ (4) & $9$ (10+) & $10$ (10+) \\
  Numerical approximations of stochastic differential equations with non-globally Lipschitz continuous coefficients & 2012 & $64$ (5) & $20$ (6) & $13$ (10+) \\
  Arbitrage and duality in nondominated discrete-time models & 2013 & $62$ (6) & $24$ (5) & $28$ (3) \\
  Existence and uniqueness theorems for solutions of McKean--Vlasov stochastic equations & 2016 & $62$ (6) & $17$ (9) & $13$ (10+) \\
  On viscosity solutions of path dependent PDEs & 2011 & $53$ (8) & $18$ (8) & $25$ (7) \\
  Backward Stochastic Differential Equations Driven by G-Brownian Motion & 2012 & $51$ (9) & $26$ (3) & $26$ (6) \\
  Functional It\^{o} calculus and stochastic integral representation of martingales & 2010 & $49$ (10) & $18$ (7) & $27$ (5) \\
  Change of variable formulas for non-anticipative functionals on path space & 2010 & $46$ (10+) & $12$ (10+) & $21$ (8) \\
  Loss of regularity for Kolmogorov equations & 2012 & $29$ (10+) & $16$ (10) & $15$ (10) \\

  A New Central Limit Theorem under Sublinear Expectations & 2008 & $28$ (10+) & $12$ (10+) & $16$ (9) \\

\hline \hline
\end{tabularx}
\end{table}

\begin{table}[ht]
\centering
\caption{Comparison of the top 10 papers in the Natural Lang. Proc., ranked by in-degree in $G$, $H$, and $I$. The table comprises the union of the top 10 lists from each network and is sorted by the in-degree in $G$, $k(G)$.}
\label{tab:rank5}
\begin{tabularx}{\textwidth}{X c ccc}
\hline \hline

  Title  & Year & $k(G)$ (Rank) & $k(H)$ (Rank) & $k(I)$ (Rank)  \\ \hline
  Attention Is All You Need & 2017 & $1\,173$ (1) &$263$ (2) & $421$ (1) \\
  Sequence to Sequence Learning with Neural Networks & 2014 & $1\,102$ (2) & $323$ (1) & $267$ (2) \\
  Distributed Representations of Words and Phrases and their Compositionality & 2013 & $567$ (3) & $120$ (3) & $183$ (3) \\
  A Call for Clarity in Reporting BLEU Scores & 2018 & $457$ (4) & $9$ (10+) & $19$ (10+) \\
  Google's Neural Machine Translation System: Bridging the Gap between Human and Machine Translation & 2016 & $332$ (5) & $81$ (4) & $52$ (5) \\
  Cross-lingual Language Model Pretraining & 2019 & $269$ (6) & $69$ (6) & $81$ (4) \\
  fairseq: A Fast, Extensible Toolkit for Sequence Modeling & 2019 & $266$ (7) & $3$ (10+) & $16$ (10+) \\
  SentencePiece: A simple and language independent subword tokenizer and detokenizer for Neural Text Processing & 2018 & $265$ (8) & $5$ (10+) & $21$ (10+) \\
  Natural Language Processing (almost) from Scratch & 2011 & $190$ (9) & $42$ (7) & $37$ (8) \\
  Multilingual Denoising Pre-training for Neural Machine Translation & 2020 & $179$ (10) & $36$ (8) & $51$ (6) \\
  Effective Approaches to Attention-based Neural Machine Translation & 2015 & $163$ (10+) & $20$ (10+) & $31$ (10) \\
  Six Challenges for Neural Machine Translation & 2014 & $135$ (10+) & $77$ (5) & $24$ (10+) \\
  Exploiting Similarities among Languages for Machine Translation & 2013 & $109$ (10+) & $35$ (9) & $30$ (10+) \\

  MASS: Masked Sequence to Sequence Pre-training for Language Generation & 2019 & $104$ (10+) & $23$ (10+) & $39$ (7) \\
  Sequence-Level Knowledge Distillation & 2016 & $88$ (10+) & $21$ (10+) & $32$ (9) \\
  Listen and Translate: A Proof of Concept for End-to-End Speech-to-Text Translation & 2016 & $78$ (10+) & $32$ (10) & $25$ (10+) \\

\hline \hline
\end{tabularx}
\end{table}

\clearpage

\section{Rank of scientists}\label{appendix_rank_scientist}

Below we report the tables with the rank of the scientists in different disciplines.

\begin{table}[ht]
\centering
\caption{Comparison of the top 10 scientist in Quantum Computing, ranked by in-degree in $G_a$, $H_a$, and $I_a$. The table comprises the union of the top 10 lists from each network and is sorted by the in-degree in $G_a$, $k(G_a)$. The names are used exactly as they appeared in the OpenAlex dataset.}
\label{tab:rank_scientist2}
\begin{tabularx}{0.7\textwidth}{l ccc}
\hline \hline

  Name  & $k(G)$ (Rank) & $k(H)$ (Rank) & $k(I)$ (Rank)  \\ \hline
  Farhi, Edward & $1\,536$ (1) &$446$ (1) & $530$ (1) \\
  Seth Lloyd & $1\,325$ (2) & $386$ (4) & $337$ (4) \\
  Gutmann, Sam & $1\,307$ (3) & $410$ (2) & $491$ (2) \\
  Goldstone, Jeffrey & $1\,193$ (4) & $392$ (3) & $466$ (3) \\
  Alán Aspuru-Guzik  & $1\,128$ (5) & $269$ (6) & $270$ (6) \\
  Charles H. Bennett & $1\,029$ (6) & $216$ (10+) & $219$ (10+) \\
  Richard Cleve & $1\,014$ (7) & $204$ (10+) & $231$ (10+) \\
  Edward Farhi & $954$ (8) & $269$ (7) & $277$ (5) \\
  Ryan Babbush  & $940$ (9) & $260$ (9) & $197$ (10+) \\
  David P. DiVincenzo & $927$ (10) & $206$ (10+) & $212$ (10+) \\
  Aram W. Harrow & $888$ (10+) & $271$ (5) & $256$ (8) \\
  Sergio Boixo & $816$ (10+) & $261$ (8) & $174$ (10+) \\
  Daniel Gottesman & $810$ (10+) & $252$ (10) & $261$ (7) \\
  Sam Gutmann & $758$ (10+) & $221$ (10+) & $245$ (10) \\
  Gottesman, Daniel & $750$ (10+) & $242$ (10+) & $255$ (9) \\

\hline \hline
\end{tabularx}
\end{table}

\begin{table}[ht]
\centering
\caption{Comparison of the top 10 scientist in Gravitational Waves, ranked by in-degree in $G_a$, $H_a$, and $I_a$. The table comprises the union of the top 10 lists from each network and is sorted by the in-degree in $G_a$, $k(G_a)$. The names are used exactly as they appeared in the OpenAlex dataset.}
\label{tab:rank_scientist3}
\begin{tabularx}{0.7\textwidth}{l ccc}
\hline \hline

  Name  & $k(G)$ (Rank) & $k(H)$ (Rank) & $k(I)$ (Rank)  \\ \hline
  Thibault Damour & $2\,707$ (1) &$433$ (1) & $549$ (1) \\
  Alessandra Buonanno & $1\,903$ (2) & $307$ (2) & $352$ (2) \\
  Masaru Shibata & $1\,385$ (3) & $199$ (5) & $190$ (6) \\
  Emanuele Berti & $1\,298$ (4) & $205$ (4) & $196$ (4) \\
  Vitor Cardoso  & $1\,258$ (5) & $218$ (3) & $210$ (3) \\
  Mark A. Scheel & $1\,218$ (6) & $168$ (8) & $176$ (8) \\
  Alessandro Nagar & $1\,172$ (7) & $162$ (10+) & $185$ (7) \\
  Luciano Rezzolla & $1\,169$ (8) & $170$ (7) & $148$ (10+) \\
  Harald P. Pfeiffer  & $1\,142$ (9) & $151$ (10+) & $150$ (10+) \\
  Bernd Brügmann & $1\,126$ (10) & $120$ (10+) & $144$ (10+) \\
  Nicolás Yunes & $1\,053$ (10+) & $168$ (9) & $194$ (5) \\
  Paolo Pani & $1\,044$ (10+) & $178$ (6) & $172$ (10) \\
  John G. Baker & $631$ (10+) & $166$ (10) & $162$ (10+) \\
  Éanna É. Flanagan & $685$ (10+) & $161$ (10+) & $176$ (9) \\  

\hline \hline
\end{tabularx}
\end{table}

\begin{table}[ht]
\centering
\caption{Comparison of the top 10 scientist in Stochastic Finance, ranked by in-degree in $G_a$, $H_a$, and $I_a$. The table comprises the union of the top 10 lists from each network and is sorted by the in-degree in $G_a$, $k(G_a)$. The names are used exactly as they appeared in the OpenAlex dataset.}
\label{tab:rank_scientist4}
\begin{tabularx}{0.7\textwidth}{l ccc}
\hline \hline

  Name  & $k(G)$ (Rank) & $k(H)$ (Rank) & $k(I)$ (Rank)  \\ \hline
  Arnulf Jentzen & $396$ (1) &$112$ (1) & $119$ (1) \\
  Martin Hutzenthaler & $240$ (2) & $71$ (2) & $74$ (2) \\
  Jentzen, Arnulf & $178$ (3) & $44$ (10) & $42$ (10+) \\
  Nizar Touzi & $159$ (4) & $45$ (8) & $55$ (4) \\
  Samy Tindel  & $159$ (5) & $48$ (6) & $45$ (10) \\
  David Nualart & $155$ (6) & $53$ (3) & $66$ (3) \\
  Peng, Shige & $150$ (7) & $52$ (5) & $52$ (5) \\
  Hu, Mingshang & $147$ (8) & $53$ (4) & $51$ (6) \\
  Ying Hu  & $145$ (9) & $47$ (7) & $46$ (9) \\
  Peter E. Kloeden & $134$ (10) & $45$ (9) & $44$ (10+) \\
  Rama Cont & $119$ (10+) & $33$ (10+) & $50$ (7) \\
  David-Antoine Fournié & $95$ (10+) & $30$ (10+) & $48$ (8)\\

\hline \hline
\end{tabularx}
\end{table}

\begin{table}[ht]
\centering
\caption{Comparison of the top 10 scientist in Natural Lang. Proc., ranked by in-degree in $G_a$, $H_a$, and $I_a$. The table comprises the union of the top 10 lists from each network and is sorted by the in-degree in $G_a$, $k(G_a)$. The names are used exactly as they appeared in the OpenAlex dataset.}
\label{tab:rank_scientist5}
\begin{tabularx}{0.7\textwidth}{l ccc}
\hline \hline

  Name  & $k(G)$ (Rank) & $k(H)$ (Rank) & $k(I)$ (Rank)  \\ \hline
  Sutskever, Ilya & $1707$ (1) &$482$ (1) & $486$ (1) \\
  Le, Quoc V. & $1494$ (2) & $447$ (2) & $360$ (10) \\
  Vinyals, Oriol & $1368$ (3) & $409$ (3) & $323$ (10+) \\
  Uszkoreit, Jakob & $1361$ (4) & $290$ (4) & $464$ (2) \\
  Parmar, Niki  & $1289$ (5) & $277$ (5) & $436$ (5) \\
  Vaswani, Ashish & $1286$ (6) & $275$ (7) & $445$ (3) \\
  Shazeer, Noam & $1277$ (7) & $277$ (6) & $438$ (4) \\
  Jones, Llion & $1220$ (8) & $267$ (9) & $425$ (7) \\
  Kaiser, Lukasz  & $1212$ (9) & $269$ (8) & $427$ (6) \\
  Gomez, Aidan N. & $1197$ (10) & $263$ (10+) & $423$ (8) \\
  Polosukhin, Illia & $1174$ (10+) & $264$ (10) & $422$ (9) \\

\hline \hline
\end{tabularx}
\end{table}

\end{document}